\documentclass[reprint,unsortedaddress,superscriptaddress,footinbib,amsmath,amssymb,aps,pra,showkeys,showpacs]{revtex4-2}
\usepackage{graphicx}
\usepackage{xcolor}
\usepackage{amsmath}
\usepackage{amssymb}
\usepackage{float}
\usepackage{braket}

\graphicspath{{./figures/}}

\newcommand{\aver}[1]{\langle #1 \rangle}
\newcommand{\epsarg}[1]{\varepsilon_{\textsc{#1}}}
\newcommand{\xiarg}[1]{\xi_{\textsc{#1}}}
\newcommand{\sxiarg}[1]{\widetilde{\xi}_{\textsc{#1}}}
\newcommand{\NOneUp}{N_{1\uparrow}}

\newcommand{\NTwoDw}{N_{2\downarrow}}
\newcommand{\dOneUp}{d_{1\uparrow}}
\newcommand{\dOneDw}{d_{1\downarrow}}
\newcommand{\dTwoUp}{d_{2\uparrow}}
\newcommand{\dTwoDw}{d_{2\downarrow}}
\newcommand{\qda}{\text{QD}_{1}}
\newcommand{\qdb}{\text{QD}_{2}}
\newcommand{\qdab}{\text{QD}_{12}}

\begin{document}

\title{A Cooper-pair beam splitter as a feasible source of entangled electrons}

\author{B. Sharmila}
\email{Sharmila.Balamurugan@warwick.ac.uk}
\affiliation{Department of Physics, University of Warwick, Coventry CV4 7AL, United Kingdom}
\author{F. M. Souza}
\email{fmsouza@ufu.br}
\affiliation{Instituto de F\'{i}sica, Universidade Federal de Uberl\^andia,
38400-902, MG, Brazil}
\author{H. M. Vasconcelos}
\email{hilma@ufc.br}
\affiliation{Departamento de Engenharia de Teleinform\'atica, Universidade Federal do Cear\'a, Fortaleza, Cear\'a, Brazil,}
\author{L. Sanz}
\email{lsanz@ufu.br}
\affiliation{Instituto de F\'{i}sica, Universidade Federal  de Uberl\^andia,
38400-902, MG, Brazil}
\date{\today}

\begin{abstract}
We investigate the generation of an entangled electron pair emerging from a system composed of two quantum dots attached to a superconductor Cooper pair beam splitter. We take into account three processes: Crossed Andreev Reflection, cotunneling, and Coulomb interaction. Together, these processes play crucial roles in the formation of entangled electronic states, with electrons being in spatially separated quantum dots. By using perturbation theory, we derive an analytical effective model that allows a simple picture of the intricate process behind the formation of the entangled state. Several entanglement quantifiers, including quantum mutual information, negativity, and concurrence, are employed to validate our findings. Finally, we define and calculate the covariance associated with the detection of two electrons, each originating from one of the quantum dots with a specific spin value. The time evolution of this observable follows the dynamics of all entanglement quantifiers, thus suggesting that it can be a useful tool for mapping the creation of entangled electrons in future applications within quantum information protocols.

\end{abstract}


\maketitle

\section{Introduction}
\label{sec:intro} 

Entanglement, as a resource, is a central key in the experimental realization of quantum computation. The generation of entangled states has recently been investigated across a range of physical systems~\cite{Friis2019}. Nevertheless, preparing entangled particles and quantifying or detecting the degree of entanglement can be challenging. Some works use the reconstruction of the density matrix to calculate fidelity with a target state~\cite{Schmied16}. In other cases, an entanglement witness is used~\cite{Terhal00} to probe the degree of entanglement. Nowadays, quantum computers capable of executing quantum algorithms rely on superconductor circuits~\cite{Castelvecchi2017}, among other possibilities~\cite{review22_40yearsQC}. The integration of any physical platform based on superconductors with other systems is strategic, since certain operations can be performed faster in those other systems. In this scenario, a hybrid architecture with superconductors and semiconductor nanostructures emerges as an interesting possibility. Several physical phenomena couple these two systems, making such integration promising. 

In a previous work, some of us investigated a Cooper-pair beam splitter, a physical system that couples a superconductor lead of Cooper pairs with two separated quantum dots~\cite{assuncao18,hofstetter2009,Weymann17}. This system has been explored recently, and important reports include the first experimental detection of spin cross correlations~\cite{bordoloi2022}. Subsequently, we present a proof of principle of how measurements of quantum transport of electrons can be used to prove an optical effect in the nanostructure: the formation of an Autler-Townes doublet. The superconductor device enables the transfer of pairs of electrons with opposite spins between two coupled quantum dots through a crossed Andreev reflection (CAR) process. An intriguing question is as follows: Are there quantum correlations between these electrons? If the answer is yes, an existing degree of entanglement can be used as a resource for applications on quantum computing. Indeed, the potential for the Andreev process to serve as entanglers was pointed out by Recher, \textit{et al.} in 2001~\cite{Recher01}. However, some questions, such as the quantum dynamics, the evolution of the degree of entanglement, and the effects of decoherence, remain unanswered. 

In this work, we propose employing a Cooper-pair beam splitter as a viable source of entangled electrons, with potential applications in quantum information and computation protocols, similar to the established utilization of entangled photons~\cite{Kwiat95,FabreRev20}. We explore the quantum dynamics of the system, modeling the crossed Andreev reflection (CAR) and cotunneling, and studying the evolution of the degree of entanglement between the two emerging electrons as they transit from the system to two connected reservoirs in the hybrid system. 

We begin with a general second-quantization Hamiltonian and first check the properties of the physical system, in order to find a two-particle subspace where the generation of an entangled state is possible. This involves constructing an effective two-particle Hamiltonian and a two-qubit model. Then, we explore the full Hamiltonian to understand the general properties of the entanglement of the eigenstates and the quantum dynamics of the system. This is achieved by defining tomographic entanglement indicators that capture the overall behavior of entanglement in the physical system of interest. Based on previous experience~\cite{Sharmila20a,Sharmila20b,Sharmila2019,Sharmila17}, we calculate several entanglement measurements along with the covariance and compare these results with the insights provided by the effective model. This allows us to propose the physical conditions necessary for producing entangled electrons.

This paper is organized as follows:  in Sec.~\ref{sec:theory}, we present our physical system and models, including the effective two-particle model and the two-qubit model. Additionally, we provide definitions for measurements and entanglement indicators to be used in our analysis. Section~\ref{sec:entanglement} is dedicated to show the results of the properties of the degree of entanglement between the electrons in our physical system, and in Sec.~\ref{sec:feasibility} we discuss the feasibility of the generation of the entangled particles in a realistic physical system. Finally, Sec.~\ref{sec:conc} presents our concluding remarks.

\section{Theoretical formalism}
\label{sec:theory}
\subsection{Physical system and Model}
\label{subsec:model}

\begin{figure}
	\includegraphics[width=0.45\textwidth]{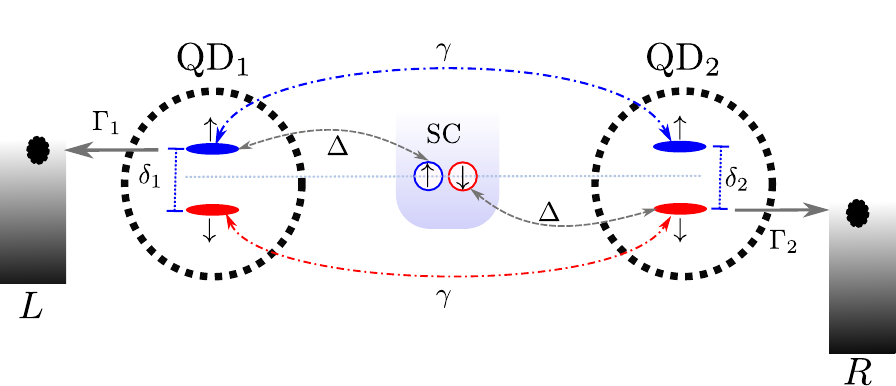}
	\caption{(Color online) Sketch of the physical system, illustrating the main parameters and interactions. Each quantum dot is represented by black dashed-line circles, with spin-up and spin-down states indicated by red and blue ellipses within each respective dot. Dotted blue lines represent Zeeman splitting with value $\delta_j$; dashed gray lines represent one of the effective CAR processes with strength $\Delta$; dot-dashed lines represent effective cotunneling processes of strength $\gamma$ for the spin-up (blue lines) and spin-down (red lines) states. The gray arrows denote the coupling with fermionic reservoirs with a rate $\Gamma_j$. The index $j=1,2$ labels the quantum dots. The Coulomb repulsion of strengths $J$ and $J'$ is not illustrated here.}
	\label{fig:model}
\end{figure}

The physical system of interest, shown in Fig.~\ref{fig:model}, consists of two quantum dots - quantum dot 1 (QD$_1$) and quantum dot 2 (QD$_{2}$) - connected through a superconductor lead. This configuration is known as a Cooper-pair beam splitter (CPBS). This kind of device creates or annihilates a pair of electrons with opposite spins at the quantum dot levels, in a process known as crossed Andreev reflection (CAR)~\cite{hiltscher2011,Trocha15}. In Fig.~\ref{fig:model}, we illustrate a scenario in which the conduction bands were occupied by an electron with spin up in QD$_1$ and spin down in QD$_2$. Zeeman splitting is given by $\delta_1$ and $\delta_2$, respectively. In contrast to our previous work~\cite{assuncao18}, we now consider cotunneling (CT) between the dots. Additionally, we take into consideration the effects of both intra- and interdot Coulomb repulsion. This combination, coupled with crossed Andreev reflection (CAR), leads to the formation of entangled states for electrons spatially separated within each quantum dot.
The modeling Hamiltonian reads as
\begin{equation}
H= H_{\textsc{z}}+H_{\textsc{c}}+ H_{\textsc{car}} + H_{\textsc{ct}},
\label{eqn:Ham}
\end{equation}
where,
\begin{eqnarray}
\label{eq:Hfullterms}
H_{\textsc{z}}&=&\sum\limits_{j=1}^{2} \tfrac{\delta_{j}}{2} \left( N_{j\uparrow} - N_{j\downarrow} \right).\\
H_{\textsc{c}}&=&J \left( N_{1\uparrow} N_{1 \downarrow} + N_{2 \uparrow} N_{2 \downarrow} \right)\nonumber\\
&&+ J' \left( N_{1\uparrow} + N_{1\downarrow}\right) \left(N_{2\uparrow} + N_{2\downarrow} \right),\nonumber\\
H_{\textsc{car}}&=&\Delta \left( \dOneUp^{\dagger} \dTwoDw^{\dagger} - \dOneDw^{\dagger} \dTwoUp^{\dagger} \right) + h.c.,\nonumber\\
H_{\textsc{ct}}&=&\gamma \left( \dOneUp^{\dagger} \dTwoUp + \dOneDw^{\dagger} \dTwoDw \right) + h.c.,\nonumber
\end{eqnarray}
Here, the diagonal terms are $H_{\textsc{z}}$, which describes the Zeeman splitting of the energy levels, given by $\delta_{j}/2$ ($j=1,2$ labeling the quantum dots in the nanostructure), and the Hamiltonian $H_{\textsc{c}}$, which accounts for intra- and inter-dot Coulomb repulsion with strengths $J$ and $J'$, respectively.
The non-diagonal terms include $H_{\textsc{car}}$, which accounts for the CAR process with strength $\Delta$, and $H_{\textsc{ct}}$, which describes cotunneling between $\qda$ and $\qdb$, associated with the parameter $\gamma$. In the equations, the operators $d_{j\sigma}$ ($d_{j\sigma}^{\dagger}$) annihilate (create) an electron with spin $\sigma$ ($\sigma=\uparrow,\downarrow$ taken along the $z$ direction) in $\text{QD}_{j}$, and $N_{j\sigma}=d_{j\sigma}^{\dagger} d_{j\sigma}$ is the number operator for quantum dot QD$_j$ and spin $\sigma$.

As indicated in [18], the Jordan-Wigner transformation proves to be a suitable method for transitioning from the second-quantization formalism to a more convenient representation for calculating physical quantities related to quantum information. Our set of four annihilation operators can be written as follows,
\begin{eqnarray}
&&d_{1\uparrow}=\sigma_{-}\otimes \mathbb{I}_{2}^{\otimes 3},\nonumber\\
&&d_{1\downarrow}=\sigma_{z}\otimes\sigma_{-}\otimes \mathbb{I}_{2}^{\otimes 2},\nonumber\\
&&d_{2\uparrow}=\sigma_{z}^{\otimes 2}\otimes\sigma_{-}\otimes \mathbb{I}_{2},\\
&&d_{2\downarrow}=\sigma_{z}^{\otimes 3}\otimes\sigma_{-},\nonumber
\end{eqnarray}
where $\sigma_{\pm}=(\sigma_{x} \pm i \sigma_{y})/2$ with $\sigma_{x}$, $\sigma_{y}$, and $\sigma_{z}$ being the standard Pauli matrices. Within this representation our model acquires a clear mapping in a $2^4$-dimensional space, with a basis given by $\{\ket{\Phi_n}\}\rightarrow\{\ket{\Phi_0}=\ket{0000}$, $\ket{\Phi_1}=\ket{0001}$, $\ket{\Phi_2}=\ket{0010}$, $\ket{\Phi_3}=\ket{0011}$, $\ket{\Phi_4}=\ket{0100}$, $\ket{\Phi_5}=\ket{0101}$, $\ket{\Phi_6}=\ket{0110}$, $\ket{\Phi_7}=\ket{0111}$, $\ket{\Phi_8}=\ket{1000}$, $\ket{\Phi_9}=\ket{1001}$, $\ket{\Phi_{10}}=\ket{1010}$, $\ket{\Phi_{11}}=\ket{1011}$, $\ket{\Phi_{12}}=\ket{1100}$, $\ket{\Phi_{13}}=\ket{1101}$, $\ket{\Phi_{14}}=\ket{1110}$, and $\ket{\Phi_{15}}=\ket{1111} \}$. It is important to notice that our model can be thought of as four coupled subsystems with two levels in each, i.e., $\ket{0}$ and $\ket{1}$. The state $\ket{0}$ ($\ket{1}$) corresponds to having one (zero) electron in the subsystem. Specifically, the index sequence in the kets follows the rule $\ket{(1-N_{1\uparrow}), (1-N_{1\downarrow}), (1-N_{2\uparrow}), (1-N_{2\downarrow})}$. This basis covers all possible combinations of occupations, spanning from the state with no particles, $\ket{1111}$ (all four levels empty), to the state with four particles, $\ket{0000}$ (all four levels occupied), in the quantum dot levels. This is particularly useful for quantum systems attached to reservoirs that can either drain all particles from the system or inject particles into the system. This approach was originally developed in the context of Lindblad operators in Ref.~\cite{Souza17}.

\subsection{The two-particle effective Model}

To investigate the physical conditions leading to the creation of entangled pairs of electrons, it is valid to project the full $2^4$-dimensional model into a reduced one composed of $2^2$ states, as previously demonstrated in~\cite{Nogueira21,Souza19,Oliveira15}. Specifically, we are interested in states resulting from the CAR processes, where each dot hosts one electron with opposite spin. To achieve this goal, we need to identify the relevant states among the $2^4$ states of the entire basis. First, we can examine the eigenvalues in the absence of the crossed Andreev reflection ($\Delta=0$) and cotunneling ($\gamma = 0$). Under these conditions, the Hamiltonian becomes diagonal, with the following $2^4$ eigenvalues:
\begin{itemize}
	\item a zero-particle state with energy $\varepsilon_{1111}=0$;
 	\item four one-particle states with energies given by $\varepsilon_{1101}=\delta_2/2$ $\varepsilon_{1110}=-\delta_2/2$, $\varepsilon_{0111}=\delta_1/2$, and $\varepsilon_{1011}=-\delta_1/2$;
 	\item six two-particle states with $\varepsilon_{0011}=\varepsilon_{1100}=J$,  $\varepsilon_{0101}=J'+\delta_1/2+\delta_2/2$, $\varepsilon_{0110}=J'+\delta_1/2-\delta_2/2$, $\varepsilon_{1001}=J'-\delta_1/2+\delta_2/2$, $\varepsilon_{1010}=J'-\delta_1/2-\delta_2/2$;
 	\item four three-particle states with $\varepsilon_{0001}=J+2J'+\delta_2/2$, $\varepsilon_{0010}=J+2J'+\delta_2/2$, $\varepsilon_{0100}=J+2J'+\delta_1/2$, $\varepsilon_{1000}=J+2J'-\delta_1/2$; and
 	\item one four-particle state with energy $\varepsilon_{0000}=2J+4J'$.
\end{itemize}
As we are looking for entangled two-electron states originating from different quantum dots, we concentrate on four of the six states within the two-particle subspace, divided into two sets: one is the pair $\left\{\ket{0110},\ket{1001}\right\}$ and the other is $\left\{\ket{1010},\ket{0101}\right\}$. From this pair of subspaces, the first subspace, consisting of the states $\ket{0110}$ and $\ket{1001}$, is a potential candidate since these states meet the requirement imposed by the superconductor lead: to put or take back a pair of electrons with two different spins from different quantum dots. A suitable set of parameters put the energies $\varepsilon_{0110}$ and $\varepsilon_{1001}$ in a resonant condition, while keeping them off resonance with other states. Assuming $J \neq 0$, $J' \neq 0$, $J \neq J'$, $\delta_1 = \delta_2 \neq 0$, we have $\varepsilon_{0110} = \varepsilon_{1001} = J'$.

The idea now is to obtain an effective two-level model, such that
\begin{equation}
H_{\mathrm{eff}} = 
\begin{bmatrix}
  \varepsilon_{0} & \Omega\\
  \Omega^* & \varepsilon_{0}
\end{bmatrix}
\end{equation}
is written in the basis $\{\ket{0110}$, $\ket{1001}\}$. To achieve this goal, we calculate the four elements of $H_{\mathrm{eff}}$ as follows:
\begin{equation}
 \bra{a} H_{\mathrm{eff}} \ket{b} = \bra{a} H_0 \ket{b} + \sum_m \frac{\bra{a} V \ket{m} \bra{m} V^\dagger \ket{b}}{E-E_m},
\end{equation}
where $H_0$ is the diagonal part of the Hamiltonian, i.e., $H_0 = H_{\textsc{c}} + H_{\textsc{z}}$, and $V$ accounts for the non-diagonal terms of the Hamiltonian, i.e., $V = H_{\textsc{car}} + H_{\textsc{ct}}$, as defined in (\ref{eq:Hfullterms}). Considering, $E \approx J'$, we find
\begin{eqnarray}
	\label{eq:omega}
\Omega&=&\bra{0110} H_{\mathrm{eff}} \ket{1001}\\
&=&- \Delta^2 \left[\frac{1}{J'}-\frac{1}{2J+3J'}\right] - \frac{2\gamma^2}{J'-J},\nonumber
\end{eqnarray}
while for the diagonal term we get
\begin{eqnarray}
 \varepsilon_0 &=& \bra{0110} H_{\mathrm{eff}} \ket{0110} \\
 &=& J' + \frac{2\gamma^2}{J'-J} + \frac{\Delta^2}{J'} - \frac{\Delta^2}{2J+3J'}.\nonumber
\end{eqnarray}
A similar expression holds for $\bra{1001} H_{\mathrm{eff}} \ket{1001}$.
With this effective model, we can describe the free quantum dynamics, i.e., not taking into account decoherence processes, according to
\begin{equation}
 \ket{\psi(t)} = e^{-\frac{i \varepsilon_0 t}{\hbar}} \left[\cos \left(\frac{\Omega t}{\hbar} \right) \ket{0110} - i \sin \left(\frac{\Omega t}{\hbar}\right) \ket{1001}\right].\label{psi_analytical}
\end{equation}
From that, we can examine, for instance, the probabilities of finding the state $\ket{0110}$ and $\ket{1001}$, $P_{0110}=\cos^2\left(\frac{\Omega t}{\hbar}\right)$
and $P_{1001}=\sin^2\left(\frac{\Omega t}{\hbar}\right)$, respectively. This analytical result will be useful for comparing and validating our numerical (exact) calculations. Additionally, it provides a characteristic timescale given by the frequency $\Omega$, an important parameter for experimental purpose in the actual generation of the entangled electrons. Our numerical results will be calculated as a function of 
\begin{equation}
	\label{eq:theta}
\theta = \Omega t,
\end{equation}
with $\hbar = 1$.
\subsection{The Two-Qubit Model}
To elucidate the origins of the entanglement properties and prepare the model for future applications on quantum computing, it is highly desirable to write a two-qubit (2QB) model. To achieve this goal, we extend the aforementioned two-particle effective model to a four-dimensional space spanned by $\{\ket{0101}, \ket{0110}, \ket{1001}, \ket{1010} \}$. Let us proceed with the following equivalence,
\begin{equation}
 \ket{01} \equiv \ket{0},  \ket{10} \equiv \ket{1},
\end{equation}
turning the basis as given to $\{\ket{00}, \ket{01}, \ket{10}, \ket{11} \}$. Notice that the states $\ket{00} \equiv \ket{0101}$ and $\ket{11} \equiv \ket{1010}$ are inert; that is, they do not couple to any other state in the present model. This can be verified through a straightforward check of the matrix representation of the full Hamiltonian. However, these states are important in our description as they facilitate the construction of a two-qubit model. With this basis extension, we can rewrite the effective model in a four-dimensional space as
\begin{equation}
H_{\mathrm{eff}}^{\mathrm{2QB}} = 
\begin{bmatrix}
  \varepsilon_{00} & 0 & 0 & 0 \\
  0 & \varepsilon_{01} & \Omega & 0 \\
  0 & \Omega & \varepsilon_{10} & 0 \\
  0 & 0 & 0 & \varepsilon_{11} \\
\end{bmatrix},
\end{equation}
with $\varepsilon_{00}\equiv \varepsilon_{0101}  = J'+\frac{\delta_{1}}{2} + \frac{\delta_{2}}{2}$, $\varepsilon_{11}\equiv \varepsilon_{1010} = J'- \frac{\delta_{1}}{2} - \frac{\delta_{2}}{2}$, and 
$\varepsilon_{01}\equiv \varepsilon_{0110} = \varepsilon_{10}\equiv \varepsilon_{1001} = \varepsilon_{0}$. Seeking clarity, it is worth noting that the Hamiltonian can be expressed as
\begin{equation}
H_{\mathrm{eff}}^{\mathrm{2QB}} = 
\begin{bmatrix}
  \delta + J' & 0 & 0 & 0 \\
  0 & \xi + J' & \Omega & 0 \\
  0 & \Omega & \xi + J' & 0 \\
  0 & 0 & 0 & -\delta + J' 
\end{bmatrix},
\end{equation}
where $\delta = \frac{\delta_{1}}{2} + \frac{\delta_{2}}{2}$ and $\xi = \frac{2\gamma^2}{J'-J} + \frac{\Delta^2}{J'} - \frac{\Delta^2}{2J + 3J'}$.
By performing an energy shift, transforming $H$ into $H - (\xi + J') I$, the Hamiltonian takes the form
\begin{equation}\label{model44}
H_{\mathrm{eff}}^{\mathrm{2QB}} = 
\begin{bmatrix}
  \delta - \xi & 0 & 0 & 0 \\
  0 & 0 & \Omega & 0 \\
  0 & \Omega & 0 & 0 \\
  0 & 0 & 0 & -\delta - \xi 
\end{bmatrix}.
\end{equation}
This makes it somewhat simpler to identify a two-qubit Hamiltonian. Let us begin by examining the off-diagonal elements. Observe that 
\begin{equation}
 \sigma_+ \otimes \sigma_- + \sigma_- \otimes \sigma_+ = 
\begin{bmatrix}\label{offdiagonalmatrix}
  0 & 0 & 0 & 0 \\
  0 & 0 & 1 & 0 \\
  0 & 1 & 0 & 0 \\
  0 & 0 & 0 & 0 
  \end{bmatrix},
 \end{equation}
provides the desired off-diagonal structure outlined in Eq.~(\ref{model44}). Using that $\sigma_+ = \frac{\sigma_x + i \sigma_y}{2}$ and $\sigma_- = \frac{\sigma_x - i \sigma_y}{2}$, we can write $\sigma_+ \otimes \sigma_- + \sigma_- \otimes \sigma_+$ $=$ $\frac{1}{2}(\sigma_x \otimes \sigma_x + \sigma_y \otimes \sigma_y)$, so that the off-diagonal part of the Hamiltonian becomes $\frac{\Omega}{2}(\sigma_x \otimes \sigma_x + \sigma_y \otimes \sigma_y)$. For the diagonal elements, we may notice that 
\begin{equation}
 \frac{1}{2}(\sigma_z \otimes \sigma_z + I \otimes I) = 
\begin{bmatrix}\label{diagonalmatrix}
  1 & 0 & 0 & 0 \\
  0 & 0 & 0 & 0 \\
  0 & 0 & 0 & 0 \\
  0 & 0 & 0 & 1 
  \end{bmatrix},
 \end{equation}
and also
\begin{equation}
 (\sigma_+ \sigma_- \otimes I + I \otimes \sigma_- \sigma_+) = 
\begin{bmatrix}
  1 & 0 & 0 & 0 \\
  0 & 0 & 0 & 0 \\
  0 & 0 & 0 & 0 \\
  0 & 0 & 0 & -1 
  \end{bmatrix}.
 \end{equation}
Combining Eqs.~(\ref{offdiagonalmatrix}) and (\ref{diagonalmatrix}) we construct the swap gate~\cite{nielsenbook}. Incorporating all these matrix structures, the effective Hamiltonian takes the form
\begin{eqnarray}
H_{\mathrm{eff}}^{\mathrm{2QB}} &=& \frac{\Omega}{2}(\sigma_x \otimes \sigma_x + \sigma_y \otimes \sigma_y) \\
                    &&-\frac{\xi}{2} (\sigma_z \otimes \sigma_z + I \otimes I) \nonumber\\
                    && +\delta (\sigma_+ \sigma_- \otimes I + I \otimes \sigma_- \sigma_+).\nonumber
\label{eq:Heff_bip_final}
\end{eqnarray}
For quantum dynamics, as the initial state is taken to be $\ket{01}=\ket{0110}$ and only the state $\ket{10}=\ket{1001}$ is accessed, we can simply assume for practical purposes that
\begin{equation}\label{eq:h2qb}
H_{\mathrm{eff}}^{\mathrm{2QB}} = \frac{\Omega}{2}(\sigma_x \otimes \sigma_x + \sigma_y \otimes \sigma_y),
\end{equation}
with $\Omega$ as defined in Eq.~(\ref{eq:omega}). This bipartite model effectively describes a system composed of two quantum dots, as illustrated in Fig.~\ref{fig:model}, with a specific focus on transitions between the states $\ket{01}=\ket{0110}$ and $\ket{10}=\ket{1001}$. With this newly derived two-qubit effective Hamiltonian, we will solve the Schrödinger equation and obtain the density matrix $\rho_{\mathrm{eff}}^{\mathrm{2QB}}$, which will be used in the next section to calculate the concurrence.

\subsection{Entanglement indicators}
In order to compute entanglement properties within the full model, described by the Hamiltonian in Eqs.~(\ref{eqn:Ham}) and (\ref{eq:Hfullterms}), and particularly relevant for an open system such as quantum dots connected to leads, we employ von Neumann entropy. This measure is defined as follows:
\begin{equation}
\xiarg{svne}^{(j)}=
-\text{Tr}\,(\rho_{j} \,\log_{2} \,\rho_{j}).
\label{eqn:SVNE}
\end{equation}
Here $\rho_{j}$ ($j=1,2$) represents the density matrix for quantum dot $\text{QD}_j$~\cite{niel}. 
In other words, let $\rho$ be the density matrix of the entire system QD$_{12}$, i.e., a 16$\times$16 matrix based on the computational basis.
The density matrix $\rho_{1}$ is obtained by taking the trace over index $k$ and $l$ of the basis $\ket{ijkl}$, specifically corresponding to the degrees of freedom of the subsystem QD$_2$. In mathematical terms, this is expressed as $\rho_{1}=\mathrm{Tr}_{2} (\rho)$. Similarly, $\rho_{2}$ is computed as $\mathrm{Tr}_{1} (\rho)$, with the trace taken over labels $i$ and $j$ of the basis. Moreover, we use the quantum mutual information ($\xiarg{qmi}$) and negativity, both being valuable metrics for quantifying quantum correlations. To be precise, the quantum mutual information is defined as:
\begin{equation}
\xiarg{qmi}=\xiarg{svne}^{(1)}+\xiarg{svne}^{(2)} - \xiarg{svne}^{(12)},
\label{eqn:QMI}
\end{equation}
where $\xiarg{svne}^{(12)} = -\text{Tr}\,(\rho \,\log_{2} \,\rho)$, while the negativity is written as
\begin{equation}
\xiarg{neg}^{(1)}=\frac{1}{2}\sum\limits_{k} \left(|\mathcal{L}_{k}|-\mathcal{L}_{k}\right).
\label{eqn:Negativity}
\end{equation}
The negativity is a measure of entanglement based on the Horodecki criterion~\cite{pptHoro}. In this context, $\lbrace\mathcal{L}_{k}\rbrace$ represents the set of eigenvalues of $\rho^{T_{1}}$, which is the partial transpose of the density matrix $\rho$ of the full system $\qdab$ with respect to the subsystem $\qda$. Equivalently, $\xiarg{neg}^{(2)}$ can be defined in terms of the  partial transpose $\rho^{T_{2}}$. For ease of notation, we will hereafter refer to $\xiarg{svne}^{(1)}$ as $\xiarg{svne}$ and $\xiarg{neg}^{(1)}$ as $\xiarg{neg}$ (where $\xiarg{neg}^{(1)}=\xiarg{neg}^{(2)}$). In the following sections, we employ scaled counterparts, denoted as $\sxiarg{qmi}=\xiarg{qmi}/2$ and $\sxiarg{neq}=2 \xiarg{neg}$, facilitating a more straightforward comparison.

We also consider a tomographic entanglement indicator referred to as $\epsarg{tei}$. To establish this indicator, we initially define the following tomographic entropies.
The bipartite tomographic entropy is given by,
\begin{align}
S_{12} = - \sum_{n} \aver{\Phi_{n}|\rho|\Phi_{n}} \log_{2} \, \aver{\Phi_{n}|\rho|\Phi_{n}}.
\label{eqn:2modeEntropy}
\end{align}
Here, $\lbrace \ket{\Phi_{n}}\rbrace$ is a convenient basis set selected to characterize the entire system $\qdab$. The tomographic entropy for the subsystem is then given by:
\begin{align}
S_{j} = - \sum_{k} \aver{\phi_{k}^{(j)}|\rho_{j}|\phi_{k}^{(j)}} \log_{2} \,\aver{\phi_{k}^{(j)}|\rho_{j}|\phi_{k}^{(j)}},
\label{eqn:1modeEntropy}
\end{align}
where $\lbrace \ket{\phi_{k}^{(j)}}\rbrace$ is the basis corresponding to the subsystem QD$_j$. Here, $\ket{\phi_{k}^{(1)}} \bra{\phi_{k}^{(1)}} = \text{Tr}_{2} \left( \ket{\Phi_{k}} \bra{\Phi_{k}} \right)$, where $\text{Tr}_{2}$ denotes the partial trace over subsystem $\qdb$. Similar expressions hold for $\lbrace \ket{\phi_{k}^{(2)}}\rbrace$.
The mutual information $\epsarg{tei}$ is expressed in terms of the tomographic entropies defined above, as:
\begin{equation}
\epsarg{tei}=S_{1} + S_{2} - S_{12}.
\label{eqn:epsTEI}
\end{equation}

In earlier studies~\cite{Sharmila20a,sharmilathesis}, it has been shown that $\epsarg{tei}$ can  exhibit signatures of entanglement, making it a valuable entanglement indicator applicable across a range of quantum systems, including continuous-variable and spin systems. We have applied a similar procedure to comprehend the entanglement behavior in the superconducting double quantum dot structure, which is the physical system of interest in the present work.

Finally, we use concurrence, as defined by Wootters~\cite{Wootters98}, as an entanglement quantifier for the 2QB model. After obtaining the solution of the Schrödinger equation for the effective 2QB Hamiltonian in Eq.~(\ref{eq:Heff_bip_final}), denoted as $\rho_{\mathrm{eff}}^{\mathrm{2QB}}$, we define an auxiliary Hermitian operator $R=\sqrt{\sqrt{\rho_{\mathrm{eff}}^{\mathrm{2QB}}}}\;\widetilde{\rho}_{\mathrm{eff}}^{\mathrm{2QB}}\sqrt{\rho_{\mathrm{eff}}^{\mathrm{2QB}}}$~\cite{Hill97}, where $\widetilde{\rho}_{\mathrm{eff}}^{\mathrm{2QB}}=(\sigma_y \otimes \sigma_y)\rho_{\mathrm{eff}}^{\mathrm{2QB}\ast}(\sigma_y \otimes \sigma_y)$ is the spin-flipped matrix, with $\rho_{\mathrm{eff}}^{\mathrm{2QB}\ast}$ being the complex conjugate of $\rho_{\mathrm{eff}}^{\mathrm{2QB}}$. The concurrence is calculated as $C(\rho_{\mathrm{eff}}^{\mathrm{2QB}})=\mathrm{max}(0,\lambda_1-\lambda_2-\lambda_3-\lambda_4)$, where $\lambda_k$ ($k=1,\ldots, 4$) represents the \emph{k}-th eigenvalue of the operator $R$ in decreasing order.

In this subsection, we have defined different entanglement indicators that are used to examine the different facets of quantum correlation. For instance, negativity explores a subset of quantum-correlated states that violate the Horodecki criterion, while indicators such as $\xiarg{qmi}$ estimate the extent of quantum correlation more generally. Comparing these two indicators as the states evolve temporally, we see that while the exact values of the indicators differ from each other, the quantitative trends conform well with each other. Further, the tomographic indicator $\epsarg{tei}$, even with a classical definition in terms of Shannon entropies, holds promise and advantage as it can be computed using projections obtained from a single basis set. We note that the set of projections corresponding to a single basis set is merely the probability distribution in that basis. It carries no information about the relative phases between these basis states, and such relative phases are crucial for a complete reconstruction of the state. This tedious reconstruction procedure is circumvented by choosing indicators such as $\epsarg{tei}$. Measures such as the concurrence, while useful in estimating the extent of entanglement, are limited solely to systems that can be effectively reduced to two qubits.

\subsection{Measurement possibilities}

To establish a connection between the preceding entanglement indicators and potential experimental detection, we assess covariances. Measurement techniques~\cite{QDmeas} frequently used in the context of quantum dots suggest that it is more convenient to \textit{estimate} the extent of entanglement through the expectation values of appropriately selected observables. For instance, covariances can be directly obtained from joint measurements of the observable $\NOneUp \NTwoDw$, making them ideal candidates in a real experiment. An interesting covariance in the context of our physical system is defined as:
\begin{equation}
\mathrm{Cov}\left(N_{1\sigma},N_{2\sigma'}\right)=4|\aver{N_{1\sigma} N_{2\sigma'}} - \aver{N_{1\sigma}} \aver{N_{2\sigma'}}|,
\label{eq:covgeneral}
\end{equation}
This quantity checks the correlation between the population of the state with spin $\sigma$ in QD$_1$ and the state with spin $\sigma'$ in QD$_2$. Additionally, we notice that the covariance is scaled by an overall factor of $4$ for ease of comparison. While not conventionally recognized as a standard entanglement indicator, the covariance in the present system provides a reasonably accurate estimate of the entanglement magnitude. The advantage of using such a \textit{non-standard} entanglement indicator is that it can be computed directly from experimental data, bypassing the need for state reconstruction.

With our analytical expression in Eq.~(\ref{psi_analytical}), we can calculate the covariance defined in Eq.~(\ref{eq:covgeneral}). For instance, we evaluate
\begin{equation}
\mathrm{Cov}\left(N_{1\uparrow},N_{2\downarrow}\right)=4|\aver{N_{1\uparrow} N_{2\downarrow}} - \aver{N_{1\uparrow}} \aver{N_{2\downarrow}}|.
\end{equation}
The first term can be written as
\begin{eqnarray}
\aver{N_{1\uparrow} N_{2\downarrow}}&
=&\left[\cos \theta \bra{0110} + i \sin \theta \bra{1001}\right].N_{1\uparrow} N_{2\downarrow}\nonumber\\ 
&&.\left[\cos \theta \ket{0110} - i \sin \theta \ket{1001}\right] \nonumber\\
&=& \cos^2 \theta.
\end{eqnarray}
Also
\begin{equation}
\aver{N_{1\uparrow}} = \aver{N_{2\downarrow}} = \cos^2 \theta,
\end{equation}
so 
\begin{equation}
\mathrm{Cov}^{\mathrm{eff}}\left(N_{1\uparrow},N_{2\downarrow}\right)= 4 \cos^2 \theta \sin^2 \theta,
\label{eqn:cov_analytical}
\end{equation}
with $\theta$ being the quantity defined in Eq.~(\ref{eq:theta}). We will compare this analytical covariance with the numerical data obtained from the complete model, as defined by Eq.~(\ref{eqn:Ham}).

\section{Entanglement properties in double quantum dot system}
\label{sec:entanglement}
\subsection{\label{subsec:eigstat}Entanglement properties and structure of the eigenstates}
\begin{figure*}
	\includegraphics[scale=0.42]{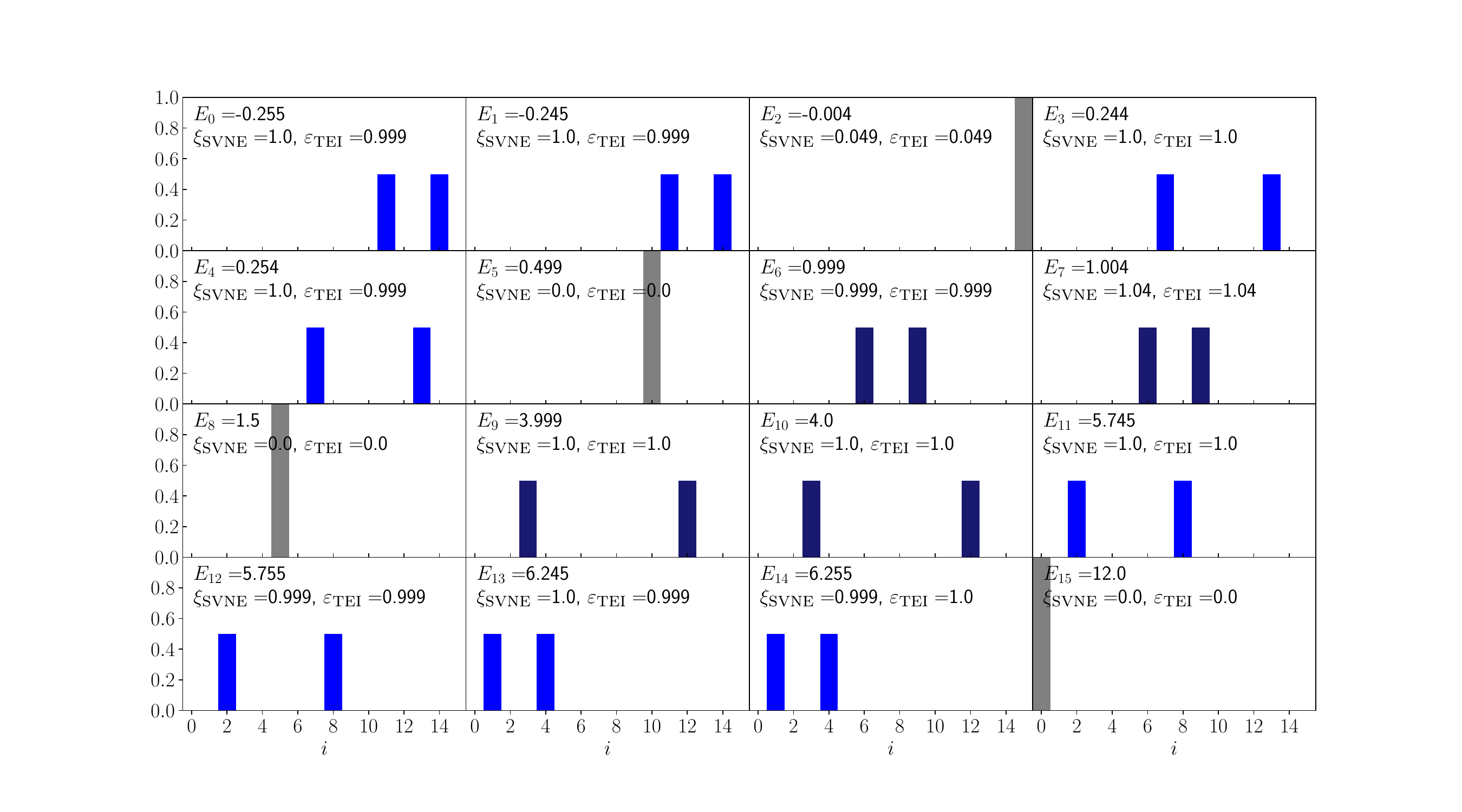}
	\caption{\label{fig:AllEigvecs} The projection $|\aver{\Phi_{i} | \psi_{n}}|^{2}$ in each of the 16 basis states $\lbrace \ket{\Phi_{i}} \rbrace$ for each $\ket{\psi_{n}}$, arranged in order of increasing energy (from lowest to highest). Label $n$ increases from above to below and left to right. Gray bars indicate eigenstates with low or no entanglement degree. The light-blue bars indicate eigenstates with high entanglement degree, but with $\mathrm{Cov}\left(N_{1\uparrow},N_{2\downarrow}\right)=0$. Finally, the dark-blue bars correspond to the two-particle entangled states, indicating eigenstates with $\mathrm{Cov}\left(N_{1\uparrow},N_{2\downarrow}\right)=1$.}
\end{figure*}

In this section, we discuss several aspects related to the properties of the set of eigenstates considering the full Hamiltonian given by Eqs.~(\ref{eqn:Ham}) and (\ref{eq:Hfullterms}). We numerically solve the Schr\"odinger equation
\begin{equation}
H \ket{\psi_{n}} = E_{n} \ket{\psi_{n}},
\end{equation}
obtaining the eigenenergies, $E_n$, and the corresponding eigenstates,  $\ket{\psi_{n}}$, labeled by $n$ (with $n=0,1,...,15$), of the Hamiltonian defined in Eq.(\ref{eqn:Ham}). In our numerical computations, we set $\Delta=0.05 J'$, $\gamma=0.005 J'$, $J=4 J'$, and $\delta_{1}=\delta_{2}=0.5 J'$ as physical parameters (see Sec.~\ref{sec:feasibility} for details). The parameters are adjusted to ensure that the levels $\varepsilon_{0110}$ and $\varepsilon_{1001}$ are virtually coupled to each other, while being energetically well separated from the other levels. This fulfills the condition that we believe is necessary to construct entangled states composed of $\ket{0110}$ and $\ket{1001}$. In Fig.~\ref{fig:AllEigvecs} we use the notation $\left\{\ket{\psi_{0}}, ..., \ket{\psi_{15}}\right\}$ for the eigenstates and 
$\left\{\ket{\Phi_{0}}, ..., \ket{\Phi_{15}}\right\}$ for the computational basis, as defined earlier.

In Fig.~\ref{fig:AllEigvecs}, we plot the projection $|\aver{\Phi_{i} | \psi_{n}}|^{2}$ in each of the 16 basis states $\lbrace \ket{\Phi_{n}} \rbrace$ for each $\ket{\psi_{n}}$, arranged in order of increasing energy (from lowest to highest). In blue, we highlight pairs of eigenstates with values of $\xiarg{svne}\sim 1$, which also have high values of $\epsarg{tei}$, a quantity mirroring the behavior of von Neumann entropy.
While the computation of $\xiarg{svne}$ for a given state requires complete knowledge of that state, the calculation of $\epsarg{tei}$ only demands the set of projections $\lbrace |\aver{\Phi_{i} | \psi_{n}}|^{2}\rbrace$, obtained for each $i$ in a specific basis set. The latter could be useful in cases where such projective measurements in the basis $\lbrace |\aver{\Phi_{i} | \psi_{n}}|^{2}\rbrace$ are experimentally feasible. 
By checking the values of the energy (refer to the label for each panel), the eigenstates are arranged in pairs, which form an anticrossing. Notice that the pairs of highly entangled eigenstates are all superpositions of two elements of the computational basis with almost equal contributions. Upon delving deeper into the analysis of the eigenstates, we can sort them into four separate sets:
\begin{enumerate}
	\item Pure states: eigenstates $\ket{\psi_2}$ (no particles), $\ket{\psi_5}$, $\ket{\psi_8}$, and $\ket{\psi_{15}}$ (four particles).
	\item One-particle entangled (with vacuum) states: eigenstates $\ket{\psi_0}$, $\ket{\psi_1}$, $\ket{\psi_3}$, and $\ket{\psi_4}$.
	\item Two-particle entangled states: eigenstates $\ket{\psi_6}$, $\ket{\psi_7}$, $\ket{\psi_9}$, and $\ket{\psi_{10}}$ (dark blue bars in Fig.~\ref{fig:AllEigvecs}).
	\item Three-particle entangled states: eigenstates $\ket{\psi_{11}}$, $\ket{\psi_{12}}$, $\ket{\psi_{13}}$, and $\ket{\psi_{14}}$.
\end{enumerate}

In our numerical and analytical analysis, we will concentrate on set (3) mentioned above. Specifically, our focus will be on the eigenstates $\ket{\psi_6}$ and $\ket{\psi_7}$, corresponding to a superposition of the basis states $\ket{6}=\ket{0110}$ and $\ket{9}=\ket{1001}$, as highlighted in dark blue in Fig.~\ref{fig:AllEigvecs}. 
In relation to the system of interest, it is valuable to examine the covariances corresponding to the four different combinations of spin states and dots.
Table~\ref{tab:coveigenstatesfig2} shows the computed values for the four quantities corresponding to each eigenstate in Fig.~\ref{fig:AllEigvecs}. Notably, the values of $\mathrm{Cov}\left(N_{1\uparrow},N_{2\downarrow}\right)$ and $\mathrm{Cov}\left(N_{1\downarrow},N_{2\uparrow}\right)$ are different from zero only for the eigenstates $\ket{\psi_{n}}$ with $n=6, \, 7, \, 9$, and $10$, as illustrated by the dark-blue bars in the figure. We observe non-zero values for $\mathrm{Cov}\left(N_{1\downarrow},N_{2\downarrow}\right)$ and $\mathrm{Cov}\left(N_{1\uparrow},N_{2\uparrow}\right)$ not only for the mentioned states $(n = 6, 7, 9, 10)$ but also for a few additional states $(n = 0, 1, 3, 4, 11, 12, 13, 14)$. Curiously, these four covariances are simultaneously equal to one for the two-particle entangled states $\ket{6}$, $\ket{7}$, $\ket{9}$ and $\ket{10}$. In the following section, we will compare the covariance with other standard entanglement quantifiers. This comparison could prove helpful in circumventing the need for detailed state reconstruction, which is typically required for computing conventional entanglement measures.

\begin{center}
\begin{widetext}
	\begin{minipage}{10.0cm}
\begin{table}[H]
	\caption{\label{tab:coveigenstatesfig2}The value of covariances $\mathrm{Cov}\left(N_{1\sigma},N_{2\sigma'}\right)$ for the eigenstates $\ket{\psi_i}$ in Fig.~\ref{fig:AllEigvecs}.}
	\begin{ruledtabular}
		\begin{tabular}{|c|c|c|c|c|}
			$i$&$\mathrm{Cov}\left(N_{1\uparrow},N_{2\downarrow}\right)$&$\mathrm{Cov}\left(N_{1\downarrow},N_{2\uparrow}\right)$&$\mathrm{Cov}\left(N_{1\downarrow},N_{2\downarrow}\right)$&$\mathrm{Cov}\left(N_{1\uparrow},N_{2\uparrow}\right)$\\
			\hline
			$0,1$&$0$&$0$&$1$&$0$\\
			\hline
			$2$&$0$&$0$&$0$&$0$\\
			\hline
			$3,4$&$0$&$0$&$0$&$1$\\
			\hline
			$5$&$0$&$0$&$0$&$0$\\
			\hline
			$6,7$&$1$&$1$&$1$&$1$\\
			\hline
			$8$&$0$&$0$&$0$&$0$\\
			\hline
			$9,10$&$1$&$1$&$1$&$1$\\
			\hline
			$11,12$&$0$&$0$&$0$&$1$\\
			\hline
			$13,14$&$0$&$0$&$1$&$0$\\
			\hline
			$15$&$0$&$0$&$0$&$0$\\
		\end{tabular}
	\end{ruledtabular}
\end{table}
\end{minipage}
\end{widetext}
\end{center}

\subsection{Entanglement dynamics in double quantum dot system}
\label{sec:dynamics}
In this section, we explore the dynamics of the system by examining the behavior of entanglement quantifiers and the covariance associated with a specific initial condition. The temporal evolution of the system is governed by the Gorini-Kossakowski-Sudarshan-Lindblad (GKSL) equation~\cite{Breuerbook}, 
\begin{align}
\nonumber\dot{\rho}(t) = -i [H,\rho(t)] &+ \Gamma_{1} \left[\dOneUp \rho(t) \dOneUp^{\dagger} - \tfrac{1}{2} \lbrace \dOneUp^{\dagger} \dOneUp, \rho(t) \rbrace \right] \\
&+ \Gamma_{2} \left[\dTwoDw \rho(t) \dTwoDw^{\dagger} - \tfrac{1}{2} \lbrace \dTwoDw^{\dagger} \dTwoDw, \rho(t) \rbrace \right],
\label{eqn:MasterEqn}
\end{align}
where $\rho(t)$ is the density matrix of the system at time $t$, and $\Gamma_j$ denotes the tunneling strength
between the quantum dots and the electronic reservoirs. In order to couple spin up electrons to the left ($\Gamma_1$) and spin down
electrons to the right ($\Gamma_2$), one can use ferromagnetic leads that provide spin-dependent coupling parameters~\cite{Dehollain20}. As single electron tunneling processes can be experimentally detected~\cite{Gustavsson06,Fujisawa_2006}, incorporating these processes into our dynamics makes our model more aligned with experimental implementations. Basically, the roles of $\Gamma_1$ and $\Gamma_2$ in the dynamics involve inducing decoherence by draining electrons from the system. This is one advantage in the use of the complete computational model composed of 16 states. In the case of the full model, the drainage of particles from the system can be accounted for when states such as $\ket{1111}$ (no particles in the system) are present in the basis.

\begin{figure}[H]
\includegraphics[width=0.5\textwidth]{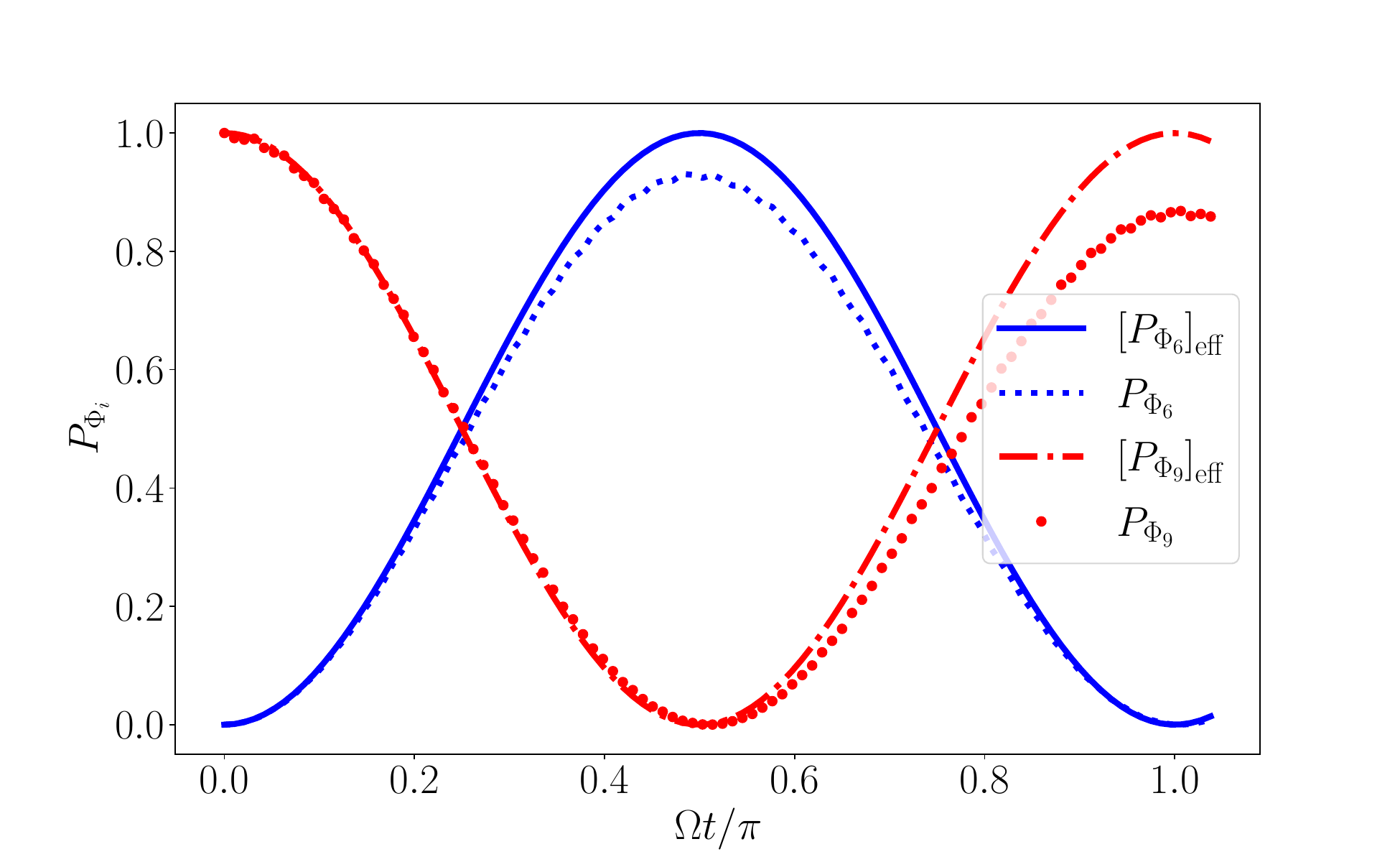}
\caption{\label{fig:OccDyn} Occupations $P_{\Phi_{i}}$ ($\left[P_{\Phi_{i}}\right]_{\mathrm{eff}}$ from the effective 2QB model) as functions of time (closed dynamics), considering the physical parameters $\Delta=0.05J'$, $\gamma=0.005J'$, $J=4J'$, $\delta_{1}=\delta_{2}=0.5J'$: $P_{\Phi_{6}}$ (solid blue line), $\left[P_{\Phi_{6}}\right]_{\mathrm{eff}}$ (dashed blue line), $P_{\Phi_{9}}$ (dot-dashed red line) and $\left[P_{\Phi_{9}}\right]_{\mathrm{eff}}$ (red dots).}
\end{figure}

We considered $\rho(0)=\ket{\Phi_{9}}\bra{\Phi_{9}}$ as the initial state, favoring the generation of entangled states $\ket{\psi_6}$ or $\ket{\psi_7}$, as illustrated in Fig.~\ref{fig:AllEigvecs}. This initialization can also be performed using ferromagnetic leads, similarly as described as Dehollain \textit{et al.}~\cite{Dehollain20}. We set the tunneling rates as $\Gamma_{1}=\Gamma_{2}=10^{-4}J'$, to simulate electrons being drained to the leads and possible being detected. Let us begin exploring the population evolution. In Fig.~\ref{fig:OccDyn} we plot the occupation probabilities $P_{\Phi_6}=\aver{\Phi_{6} | \rho(t) | \Phi_{6}}$ and $P_{\Phi_9}=\aver{\Phi_{9} | \rho(t) | \Phi_{9}}$. We also display $[P_{\Phi_6}]_{\mathrm{eff}}$ and $[P_{\Phi_9}]_{\mathrm{eff}}$, which represent the occupation probabilities of the states $\ket{\Phi_6}$ and $\ket{\Phi_9}$ obtained from the effective model, as given by Eq. (\ref{psi_analytical}). The adopted timescale is determined by $\theta=\Omega t$ ($\hbar=1$). We observe that at instants $\theta = \pi/4$ and $3 \pi /4$, the time-evolved state has nearly equal contributions from the basis states 
$\ket{\Phi_{6}}$ and $\ket{\Phi_{9}}$, with $P_{\Phi_6}$=$P_{\Phi_9}$ $\approx$ $1/2$. In particular, at $\theta = 3 \pi/4$, we can already observe the effects of decoherence induced by the tunnel coupling between dots and leads, leading to a slight suppression of the oscillation amplitudes. The populations $[P_{\Phi_6}]_{\mathrm{eff}}$ and $[P_{\Phi_9}]_{\mathrm{eff}}$ do not exhibit damping, since the effective model does not consider coupling to the leads. The populations for the other states $\ket{\Phi_i}$ remain close to zero throughout the dynamics, consistent with our parameters set that favors the occupation of only $\ket{\Phi_6}$ and $\ket{\Phi_9}$.

\begin{figure}[h]
\includegraphics[width=0.5\textwidth]{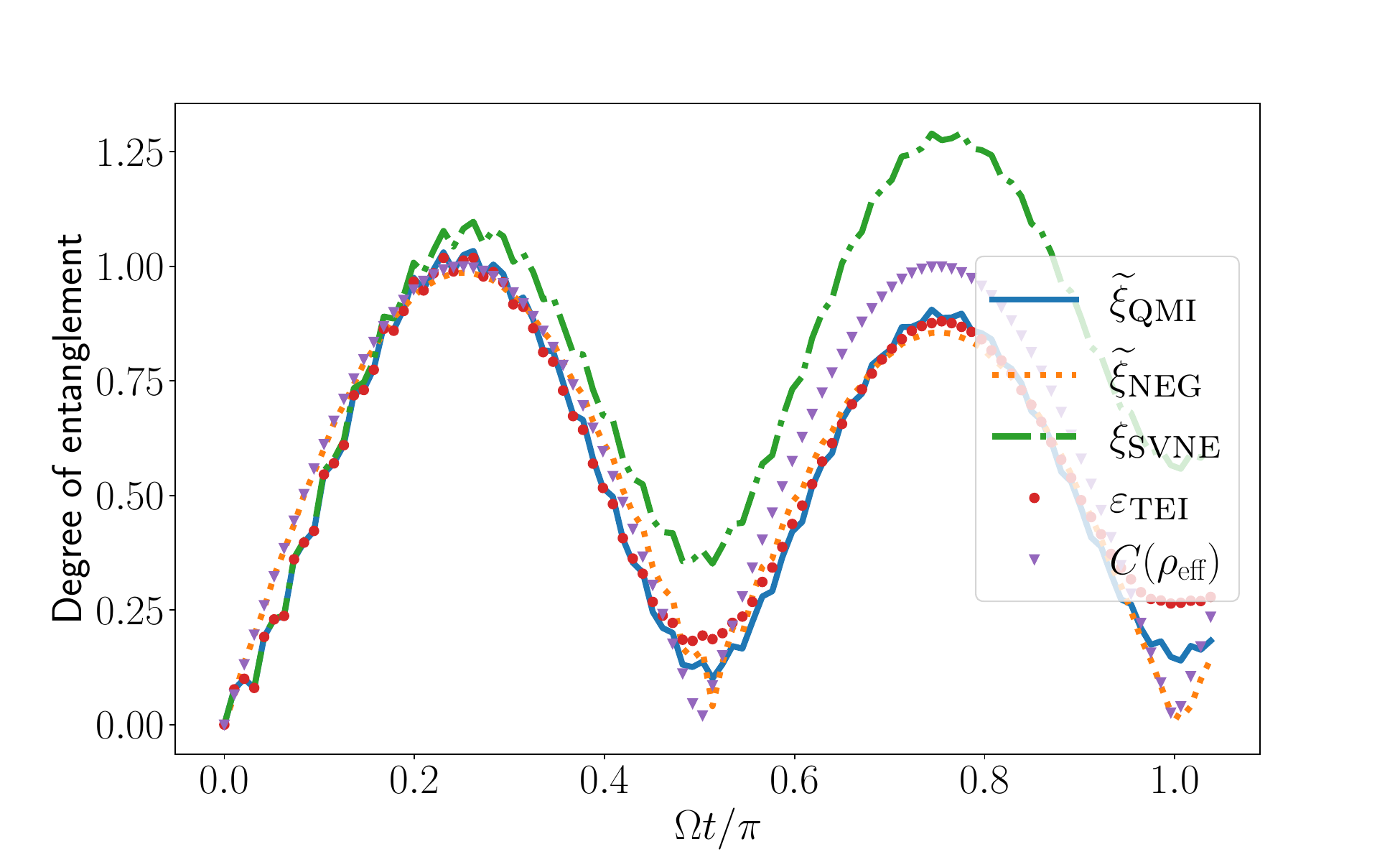}
\caption{\label{fig:EntangIndics} The temporal dynamics of some entanglement quantifiers are considered using the same physical parameters as shown in Fig.~\ref{fig:OccDyn}. Here, we illustrate $\sxiarg{qmi}=0.5 \xiarg{qmi}$ (solid blue line), $\sxiarg{neg}=2 \xiarg{neg}$ (dotted orange line), $\xiarg{svne}$ (dot-dashed green line), $\epsarg{tei}$ (red dots), and the concurrence $C(\rho_{\mathrm{eff}})$ (purple triangles) obtained from the effective 2QB model, where $\rho_{\mathrm{eff}}=\ket{\psi(t)}\bra{\psi(t)}$ is calculated using Eq. (\ref{psi_analytical}).}
\end{figure}

Now that we have confirmed, through the analysis of population dynamics, that the state evolves to a superposition of basis elements $\ket{\Phi_6}$ and $\ket{\Phi_9}$, it is time to examine the evolution of the entanglement quantifiers. The degree of entanglement between $\qda$ and $\qdb$ is shown in Fig.~\ref{fig:EntangIndics}, where the entanglement indicators $\sxiarg{qmi}$, $\sxiarg{neg}$, $\xiarg{svne}$ and $\epsarg{tei}$ are plotted as functions of time. These indicators are computed based on the complete model defined by Eq. (\ref{eqn:Ham}). It is important to point out that the subsystem QD$_{1}$ has a Hilbert space dimension $d=4$ as described in Sec. \ref{subsec:model} and therefore, the von Neumann entropy for that subsystem, $\xiarg{svne} \leqslant \log_{2} d$, i.e., $\xiarg{svne} \leqslant 2$. We also note that $\xiarg{svne}$ is not a suitable entanglement indicator when the state of the full bipartite system is mixed. This arises from the fact that, unlike in the case of the bipartite pure states, the eigenvalues of the reduced density matrix corresponding to either subsystem (used to compute $\xiarg{svne}$) are not longer related to entanglement measures such as the Schmidt rank. In the case of the bipartite mixed states, the von Neumann entropies corresponding to the two subsystems are not even equal to each other. For instance, $\xiarg{svne}$ corresponding to QD$_{1}$ is not equal to that corresponding to QD$_{2}$. On the other hand, $\xiarg{qmi}$ is more appropriately \textit{normalized} accounting for the mixedness of the bipartite state and is better suited as an entanglement indicator. This is evident in Fig.~\ref{fig:EntangIndics}. $\xiarg{svne}$ does not follow the same trend as the other entanglement indicators, as it captures not only the extent of quantum correlation but also how the system progressively decoheres with time. Additionally, we calculate the concurrence $C(\rho^{2QB}_{\mathrm{eff}})$ using the effective 2QB model in Eq. (\ref{eq:h2qb}). The concurrence fully agrees with the results of the full model, except for the decoherence introduced by the leads, which is not accounted for in the 2QB model calculations. The evolution of all these quantifiers follows the dynamics of the populations, thus reaching a maximum at $\theta = \pi/4$ and $3 \pi /4$, as expected. This demonstrates that the interplay between Crossed Andreev Reflection and Coulomb repulsion in the present double-dot structure can generate entangled states by properly selecting parameters to ensure that the relevant eigenstates $\ket{\psi_6}$ and $\ket{\psi_7}$ (refer to Fig. ~\ref{fig:AllEigvecs}), are energetically isolated from the other states.

\begin{figure}[h]
\includegraphics[width=0.5\textwidth]{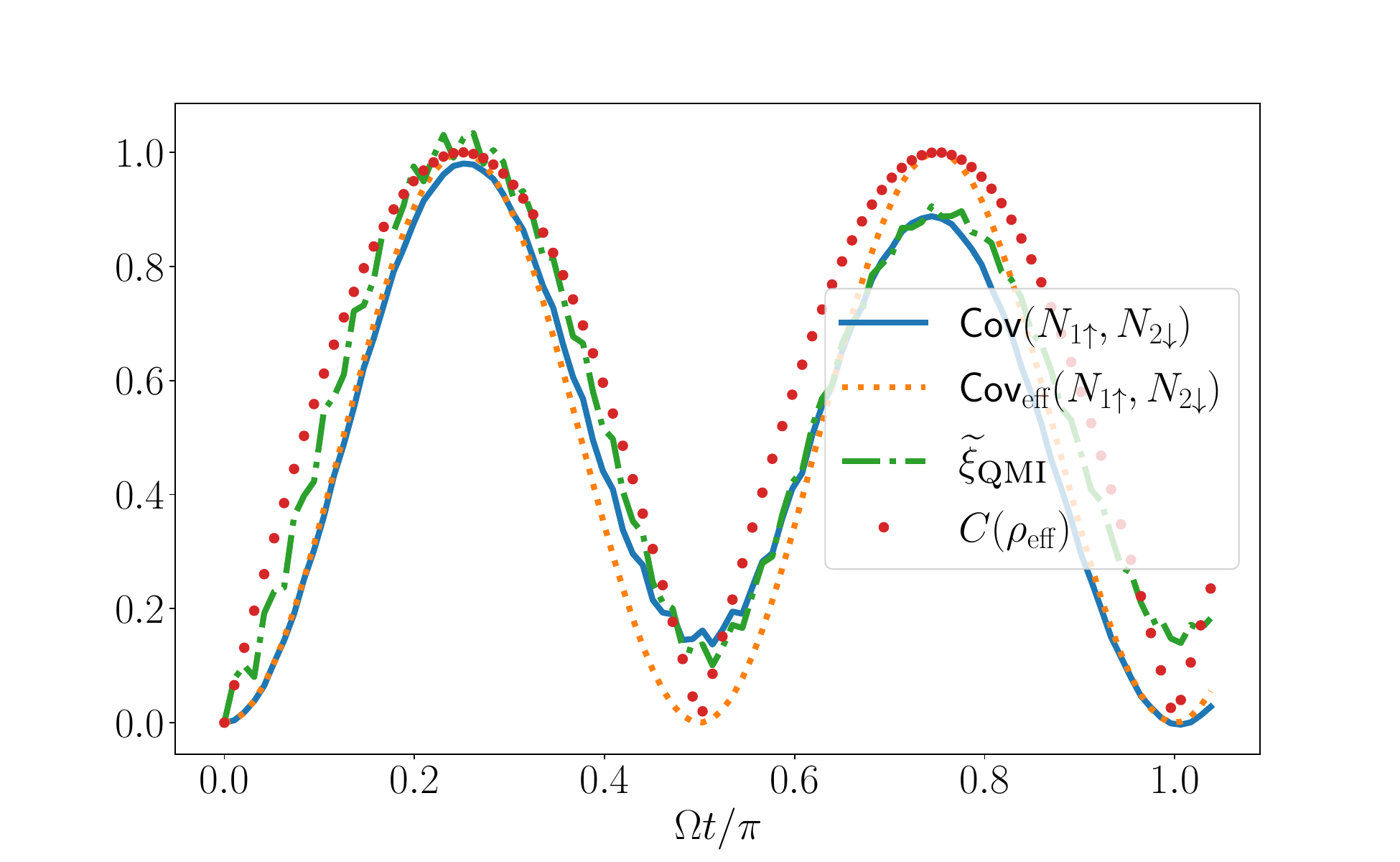}
\caption{\label{fig:Covar} Comparing the temporal evolution of the covariance $\mathrm{Cov}\left(N_{1\uparrow},N_{2\downarrow}\right)$ (blue solid line) with the corresponding covariance from the effective 2QB  model, $\mathrm{Cov}_{\mathrm{eff}}\left(N_{1\uparrow},N_{2\downarrow}\right)$ (dotted orange line), using Eq. (\ref{eqn:cov_analytical}). The scaled quantum mutual information $\sxiarg{qmi}$ (dot-dashed green line) and concurrence $C(\rho_{\mathrm{eff}})$ (red dots) are also presented, utilizing the same physical parameters as in Fig.~\ref{fig:OccDyn}.}
\end{figure}

Finally, we explore the dynamics of covariance to assess whether it can indicate the reported entanglement formation.
In Fig.~\ref{fig:Covar}, we present both $\mathrm{Cov}\left(N_{1\uparrow},N_{2\downarrow}\right)$ and $\mathrm{Cov}_{\mathrm{eff}}\left(N_{1\uparrow},N_{2\downarrow}\right)$ as a function of $\theta/\pi$. Here, $\mathrm{Cov}$ represents the covariance obtained with the full model using Eq. (\ref{eqn:Ham}), while $\mathrm{Cov}_{\mathrm{eff}}$ corresponds to the effective model given by Eq. (\ref{psi_analytical}). More specifically, for the effective model, we use the covariance provided by Eq. (\ref{eqn:cov_analytical}).
Interestingly, the covariances peak at $\theta = \pi/4$ and $3 \pi /4$, in agreement with the entanglement quantifiers presented in Fig.~\ref{fig:EntangIndics}.
To facilitate comparison in Fig.~\ref{fig:Covar}, we also include $\sxiarg{qmi}$ and $C(\rho_{\mathrm{eff}})$. This result suggests that, although covariance is not a standard entanglement quantifier, it can serve as a simple tool to indicate the potential formation of entanglement in preliminary experimental detection. Naturally, more sophisticated entanglement quantifiers or tomography are required to confirm the formation of entangled states.

\section{Experimental Feasibility}
\label{sec:feasibility}
In this section, we briefly discuss the experimental feasibility of our proposal. First, we discuss our choices of values of the physical parameters. Notice that in our previous sections, we set $J=4J'$, $\Delta=0.05J'$, $\gamma=0.005J'$, and $\Gamma_1=\Gamma_2=10^{-4}J'$. In actual experimental setups, the typical values of Coulomb interaction ranges for charged quantum dots vary from dozens to hundreds of $\mu$eV~\cite{Shinkai09,Fujisawa11}, with the intradot $J$ being stronger than the interdot Coulomb repulsion $J'$~\cite{Shinkai09,Fujisawa23}, i.e., around two to five times. In transport experiments, the coupling parameter with the reservoirs is highly tunable, i.e., adjustable via gate voltages~\cite{Gramich17}.

Concerning the tunneling rate $\Delta$, the values reported in the literature differ depending on the experimental setup. For example, the experimental works of Maisi \textit{et al}.~\cite{Maisi11,Maisi14} report tunneling between superconductor and quantum dots in the range of $10^{-6}$ $\mu$eV, while in the recent work of Bordoloi \textit{et al}.~\cite{bordoloi2022}, the values of tunneling rates vary between $59$~$\mu$eV and $130$~$\mu$eV. By selecting $J'=100$~$\mu$eV,  our proposal requires that $\Delta\approx 5$~$\mu$eV, a value that falls between those found in the literature. This tunneling process is dominant, as reported in Ref.~\cite{bordoloi2022} (around 85\% of the fraction in the conductance signal), if compared with other processes, including cotunneling. So it is reasonable to our assumption of $\gamma$ being smaller than $\Delta$.

With our choice of values of the physical parameters, it is straightforward to calculate the value of our two-particle effective model frequency, which results on $\Omega \approx 0.002 J'$. Considering again $J'=100$~$\mu$eV, we find $\Omega \approx 0.23$~$\mu$eV, which, in frequency, reads as $0.34$~GHz. This timescale is important to the following step in our analysis: the effects of decoherence. The main decoherence process in our physical system is the dephasing due to background noise in the semiconductor nanostructure forming the quantum dot system, originated by several processes. Since $\Omega$ gives the characteristic frequency of formation of the entangled state in the present dynamics, we must expect $\Omega > \Gamma_{\mathrm{deph}}$, with $\Gamma_{\mathrm{deph}}$ being the dephasing rate.
This $\Gamma_{\mathrm{deph}}$ can be as high as $1$~GHz~\cite{Fujisawa11} in GaAs quantum dots, which could compromise the formation of entanglement in the present system. However, our time scaling can be adjusted by interdot Coulomb repulsion, which depends on the physical characteristics of the nanostructure~\cite{Goldhaber98}. By raising $J'$, one can increase the characteristic frequency $\Omega$ to values close to $\Gamma_{\mathrm{deph}}$.

\begin{figure}[h]
\includegraphics[width=0.4\textwidth]{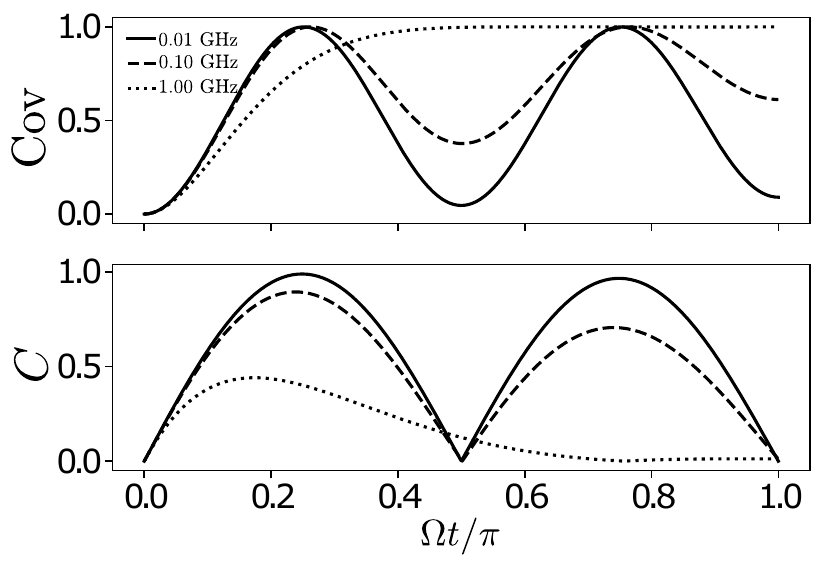}
\caption{The temporal dynamics of the Covariance and the concurrence considering the action of the dephasing considering $0.01$ GHz (solid line), $0.1$ GHz (dashed line) and $1.0$ GHz (dotted line).}
\label{fig:dephasing}
\end{figure}

In order to estimate the effects of dephasing in the present quantum dynamics, we add dephasing channels
to the Lindblad equation for the effective model, i.e.,
\begin{eqnarray}
\nonumber{\dot{\rho}_{\mathrm{eff}}^{\mathrm{2QB}}}(t) &=& -i [H_{\mathrm{eff}}^{\mathrm{2QB}},\rho_{\mathrm{eff}}^{\mathrm{2QB}}(t)] 
+\frac{1}{2}\Gamma_{\mathrm{deph}}\sum_{l=1}^2 [2 S_l \rho_{\mathrm{eff}}^{\mathrm{2QB}}(t) S_l^\dagger
\\ &&  - S_l^\dagger S_l \rho_{\mathrm{eff}}^{\mathrm{2QB}}(t) - \rho_{\mathrm{eff}}^{\mathrm{2QB}}(t) S_l^\dagger S_l ],
\end{eqnarray}
where $S_1 = \ket{01}\bra{01}$ and $S_2=\ket{10}\bra{10}$. In Fig.~\ref{fig:dephasing} we show both, the covariance ($\mathrm{Cov}$, upper panel) and the concurrence ($\mathrm{C}$, lower panel) in the presence of three values for dephasing rates, $0.01$~GHz (solid line), $0.1$~GHz (dashed line) and $1.0$~GHz (dotted line). For $\Omega \gg \Gamma_{\mathrm{deph}}$ (solid line) both $\mathrm{Cov}$ and $\mathrm{C}$ oscillate as in the case without dephasing, as compared with Fig~\ref{fig:Covar}. When $\Omega$ becomes of the same order as $\Gamma_{\mathrm{deph}}$ (dashed line), the dephasing starts to play a more significant role in the quantum evolution. Still, while the $\mathrm{Cov}$ presents peaks near one, $\mathrm{C}$ becomes close to $0.9$ in the first peak and $0.7$ in the second peak, indicating high values of entanglement as a resource for possible applications in quantum information~\cite{HorodeckiRev}. 
Finally, for $\Omega \ll \Gamma_{\mathrm{deph}}$ (dashed line) the peaks
in both $\mathrm{Cov}$ and $\mathrm{C}$ are strongly suppressed, indicating that no entangled state is being formed as time evolves. Curiously, the covariance tends to one for increasing time, due to the formation of a mixed density matrix such as $(\ket{01}\bra{01}+\ket{10}\bra{10})/2$.
So, in conclusion, in order to observe the formation of entanglement in the present setup, a dephasing rate around $0.1$ GHz would be desirable.
\section{Concluding remarks}
\label{sec:conc}
In this work, we investigate the entanglement properties of electrons emerging from a superconductor beam splitter. We aim to observe the dynamical generation of maximally entangled electronic states, where pairs of electrons populate spatially separated quantum dots. The model incorporates crossed Andreev reflection, cotunneling and electrostatic Coulomb interaction between particles within a pair of quantum dots. Additionally, the system is coupled to spin-sensitive reservoirs, enabling tunnel current measurements. We demonstrate that the eigenstates of the Hamiltonian split into non-entangled states and six subspaces of entangled states. Among the six, we focus our analysis on one of the two-particle subspaces. We show that, by appropriately adjusting the system parameters and selecting the initial state, the system evolves into a highly entangled state at times corresponding to $\theta=\pi/4$ and $\theta=3\pi/4$. In addition to conventional entanglement and correlation measurements, we demonstrate how the covariance $\mathrm{Cov}\left(N_{1\uparrow},N_{2\downarrow}\right)$ can serve as a useful tool to indicate the formation of entangled states in the present system. We also discuss the experimental feasibility of our proposal, using realistic physical parameters in our simulations and by taking into account the action of the dephasing, the main source of decoherence of the physical platform. Our results show that it is desirable to keep a dephasing rate around $0.1$ GHz in order to use this system as an actual source of entangled electrons.
\begin{acknowledgments}
F. M. Souza, H. M. Vasconcelos and L. Sanz thank CNPq for financial support (No. 422350/2021-4). B. Sharmilla acknowledges the UK STFC "Quantum Technologies for Fundamental Physics" program (Grant no. ST/T006404/1) for support.
\end{acknowledgments}


\begin{thebibliography}{41}%
	\makeatletter
	\providecommand \@ifxundefined [1]{%
		\@ifx{#1\undefined}
	}%
	\providecommand \@ifnum [1]{%
		\ifnum #1\expandafter \@firstoftwo
		\else \expandafter \@secondoftwo
		\fi
	}%
	\providecommand \@ifx [1]{%
		\ifx #1\expandafter \@firstoftwo
		\else \expandafter \@secondoftwo
		\fi
	}%
	\providecommand \natexlab [1]{#1}%
	\providecommand \enquote  [1]{``#1''}%
	\providecommand \bibnamefont  [1]{#1}%
	\providecommand \bibfnamefont [1]{#1}%
	\providecommand \citenamefont [1]{#1}%
	\providecommand \href@noop [0]{\@secondoftwo}%
	\providecommand \href [0]{\begingroup \@sanitize@url \@href}%
	\providecommand \@href[1]{\@@startlink{#1}\@@href}%
	\providecommand \@@href[1]{\endgroup#1\@@endlink}%
	\providecommand \@sanitize@url [0]{\catcode `\\12\catcode `\$12\catcode `\&12\catcode `\#12\catcode `\^12\catcode `\_12\catcode `\%12\relax}%
	\providecommand \@@startlink[1]{}%
	\providecommand \@@endlink[0]{}%
	\providecommand \url  [0]{\begingroup\@sanitize@url \@url }%
	\providecommand \@url [1]{\endgroup\@href {#1}{\urlprefix }}%
	\providecommand \urlprefix  [0]{URL }%
	\providecommand \Eprint [0]{\href }%
	\providecommand \doibase [0]{https://doi.org/}%
	\providecommand \selectlanguage [0]{\@gobble}%
	\providecommand \bibinfo  [0]{\@secondoftwo}%
	\providecommand \bibfield  [0]{\@secondoftwo}%
	\providecommand \translation [1]{[#1]}%
	\providecommand \BibitemOpen [0]{}%
	\providecommand \bibitemStop [0]{}%
	\providecommand \bibitemNoStop [0]{.\EOS\space}%
	\providecommand \EOS [0]{\spacefactor3000\relax}%
	\providecommand \BibitemShut  [1]{\csname bibitem#1\endcsname}%
	\let\auto@bib@innerbib\@empty
	\bibitem [{\citenamefont {Friis}\ \emph {et~al.}(2019)\citenamefont {Friis}, \citenamefont {Vitagliano}, \citenamefont {Malik},\ and\ \citenamefont {Huber}}]{Friis2019}%
	\BibitemOpen
	\bibfield  {author} {\bibinfo {author} {\bibfnamefont {N.}~\bibnamefont {Friis}}, \bibinfo {author} {\bibfnamefont {G.}~\bibnamefont {Vitagliano}}, \bibinfo {author} {\bibfnamefont {M.}~\bibnamefont {Malik}},\ and\ \bibinfo {author} {\bibfnamefont {M.}~\bibnamefont {Huber}},\ }\bibfield  {title} {\bibinfo {title} {Entanglement certification from theory to experiment},\ }\href {https://doi.org/10.1038/s42254-018-0003-5} {\bibfield  {journal} {\bibinfo  {journal} {Nature Reviews Physics}\ }\textbf {\bibinfo {volume} {1}},\ \bibinfo {pages} {72} (\bibinfo {year} {2019})}\BibitemShut {NoStop}%
	\bibitem [{\citenamefont {Schmied}(2016)}]{Schmied16}%
	\BibitemOpen
	\bibfield  {author} {\bibinfo {author} {\bibfnamefont {R.}~\bibnamefont {Schmied}},\ }\bibfield  {title} {\bibinfo {title} {Quantum state tomography of a single qubit: comparison of methods},\ }\href {https://doi.org/10.1080/09500340.2016.1142018} {\bibfield  {journal} {\bibinfo  {journal} {Journal of Modern Optics}\ }\textbf {\bibinfo {volume} {63}},\ \bibinfo {pages} {1744} (\bibinfo {year} {2016})},\ \Eprint {https://arxiv.org/abs/https://doi.org/10.1080/09500340.2016.1142018} {https://doi.org/10.1080/09500340.2016.1142018} \BibitemShut {NoStop}%
	\bibitem [{\citenamefont {Terhal}(2000)}]{Terhal00}%
	\BibitemOpen
	\bibfield  {author} {\bibinfo {author} {\bibfnamefont {B.~M.}\ \bibnamefont {Terhal}},\ }\bibfield  {title} {\bibinfo {title} {Bell inequalities and the separability criterion},\ }\href {https://doi.org/https://doi.org/10.1016/S0375-9601(00)00401-1} {\bibfield  {journal} {\bibinfo  {journal} {Physics Letters A}\ }\textbf {\bibinfo {volume} {271}},\ \bibinfo {pages} {319} (\bibinfo {year} {2000})}\BibitemShut {NoStop}%
	\bibitem [{\citenamefont {Castelvecchi}(2017)}]{Castelvecchi2017}%
	\BibitemOpen
	\bibfield  {author} {\bibinfo {author} {\bibfnamefont {D.}~\bibnamefont {Castelvecchi}},\ }\bibfield  {title} {\bibinfo {title} {Quantum computers ready to leap out of the lab in 2017},\ }\href {https://doi.org/10.1038/541009a} {\bibfield  {journal} {\bibinfo  {journal} {Nature}\ }\textbf {\bibinfo {volume} {541}},\ \bibinfo {pages} {9} (\bibinfo {year} {2017})}\BibitemShut {NoStop}%
	\bibitem [{rev(2022)}]{review22_40yearsQC}%
	\BibitemOpen
	\bibfield  {title} {\bibinfo {title} {40 years of quantum computing},\ }\href {https://doi.org/10.1038/s42254-021-00410-6} {\bibfield  {journal} {\bibinfo  {journal} {Nature Reviews Physics}\ }\textbf {\bibinfo {volume} {4}},\ \bibinfo {pages} {1} (\bibinfo {year} {2022})}\BibitemShut {NoStop}%
	\bibitem [{\citenamefont {Assunção}\ \emph {et~al.}(2018)\citenamefont {Assunção}, \citenamefont {Diniz}, \citenamefont {Sanz},\ and\ \citenamefont {Souza}}]{assuncao18}%
	\BibitemOpen
	\bibfield  {author} {\bibinfo {author} {\bibfnamefont {M.~O.}\ \bibnamefont {Assunção}}, \bibinfo {author} {\bibfnamefont {G.~S.}\ \bibnamefont {Diniz}}, \bibinfo {author} {\bibfnamefont {L.}~\bibnamefont {Sanz}},\ and\ \bibinfo {author} {\bibfnamefont {F.~M.}\ \bibnamefont {Souza}},\ }\bibfield  {title} {\bibinfo {title} {Autler-townes doublet observation via a cooper-pair beam splitter},\ }\href {https://doi.org/10.1103/PhysRevB.98.075423} {\bibfield  {journal} {\bibinfo  {journal} {Phys. Rev. B}\ }\textbf {\bibinfo {volume} {98}},\ \bibinfo {pages} {075423} (\bibinfo {year} {2018})}\BibitemShut {NoStop}%
	\bibitem [{\citenamefont {Hofstetter}\ \emph {et~al.}(2009)\citenamefont {Hofstetter}, \citenamefont {Csonka}, \citenamefont {Nygard},\ and\ \citenamefont {Sch\"onenberger}}]{hofstetter2009}%
	\BibitemOpen
	\bibfield  {author} {\bibinfo {author} {\bibfnamefont {L.}~\bibnamefont {Hofstetter}}, \bibinfo {author} {\bibfnamefont {S.}~\bibnamefont {Csonka}}, \bibinfo {author} {\bibfnamefont {J.}~\bibnamefont {Nygard}},\ and\ \bibinfo {author} {\bibfnamefont {C.}~\bibnamefont {Sch\"onenberger}},\ }\bibfield  {title} {\bibinfo {title} {Cooper pair splitter realized in a two-quantum-dot y-junction},\ }\href@noop {} {\bibfield  {journal} {\bibinfo  {journal} {Nature}\ }\textbf {\bibinfo {volume} {461}},\ \bibinfo {pages} {960} (\bibinfo {year} {2009})}\BibitemShut {NoStop}%
	\bibitem [{\citenamefont {Wrze\ifmmode~\acute{s}\else \'{s}\fi{}niewski}\ and\ \citenamefont {Weymann}(2017)}]{Weymann17}%
	\BibitemOpen
	\bibfield  {author} {\bibinfo {author} {\bibfnamefont {K.}~\bibnamefont {Wrze\ifmmode~\acute{s}\else \'{s}\fi{}niewski}}\ and\ \bibinfo {author} {\bibfnamefont {I.}~\bibnamefont {Weymann}},\ }\bibfield  {title} {\bibinfo {title} {Kondo physics in double quantum dot based cooper pair splitters},\ }\href {https://doi.org/10.1103/PhysRevB.96.195409} {\bibfield  {journal} {\bibinfo  {journal} {Phys. Rev. B}\ }\textbf {\bibinfo {volume} {96}},\ \bibinfo {pages} {195409} (\bibinfo {year} {2017})}\BibitemShut {NoStop}%
	\bibitem [{\citenamefont {Bordoloi}\ \emph {et~al.}(2022)\citenamefont {Bordoloi}, \citenamefont {Zannier}, \citenamefont {Sorba}, \citenamefont {Sch\"onenberger},\ and\ \citenamefont {Baumgartner}}]{bordoloi2022}%
	\BibitemOpen
	\bibfield  {author} {\bibinfo {author} {\bibfnamefont {L.}~\bibnamefont {Bordoloi}}, \bibinfo {author} {\bibfnamefont {V.}~\bibnamefont {Zannier}}, \bibinfo {author} {\bibfnamefont {L.}~\bibnamefont {Sorba}}, \bibinfo {author} {\bibfnamefont {C.}~\bibnamefont {Sch\"onenberger}},\ and\ \bibinfo {author} {\bibfnamefont {A.}~\bibnamefont {Baumgartner}},\ }\bibfield  {title} {\bibinfo {title} {Spin cross-correlation experiments in an electron entangler},\ }\href@noop {} {\bibfield  {journal} {\bibinfo  {journal} {Nature}\ }\textbf {\bibinfo {volume} {612}},\ \bibinfo {pages} {454} (\bibinfo {year} {2022})}\BibitemShut {NoStop}%
	\bibitem [{\citenamefont {Recher}\ \emph {et~al.}(2001)\citenamefont {Recher}, \citenamefont {Sukhorukov},\ and\ \citenamefont {Loss}}]{Recher01}%
	\BibitemOpen
	\bibfield  {author} {\bibinfo {author} {\bibfnamefont {P.}~\bibnamefont {Recher}}, \bibinfo {author} {\bibfnamefont {E.~V.}\ \bibnamefont {Sukhorukov}},\ and\ \bibinfo {author} {\bibfnamefont {D.}~\bibnamefont {Loss}},\ }\bibfield  {title} {\bibinfo {title} {Andreev tunneling, coulomb blockade, and resonant transport of nonlocal spin-entangled electrons},\ }\href {https://doi.org/10.1103/PhysRevB.63.165314} {\bibfield  {journal} {\bibinfo  {journal} {Phys. Rev. B}\ }\textbf {\bibinfo {volume} {63}},\ \bibinfo {pages} {165314} (\bibinfo {year} {2001})}\BibitemShut {NoStop}%
	\bibitem [{\citenamefont {Kwiat}\ \emph {et~al.}(1995)\citenamefont {Kwiat}, \citenamefont {Mattle}, \citenamefont {Weinfurter}, \citenamefont {Zeilinger}, \citenamefont {Sergienko},\ and\ \citenamefont {Shih}}]{Kwiat95}%
	\BibitemOpen
	\bibfield  {author} {\bibinfo {author} {\bibfnamefont {P.~G.}\ \bibnamefont {Kwiat}}, \bibinfo {author} {\bibfnamefont {K.}~\bibnamefont {Mattle}}, \bibinfo {author} {\bibfnamefont {H.}~\bibnamefont {Weinfurter}}, \bibinfo {author} {\bibfnamefont {A.}~\bibnamefont {Zeilinger}}, \bibinfo {author} {\bibfnamefont {A.~V.}\ \bibnamefont {Sergienko}},\ and\ \bibinfo {author} {\bibfnamefont {Y.}~\bibnamefont {Shih}},\ }\bibfield  {title} {\bibinfo {title} {New high-intensity source of polarization-entangled photon pairs},\ }\href {https://doi.org/10.1103/PhysRevLett.75.4337} {\bibfield  {journal} {\bibinfo  {journal} {Phys. Rev. Lett.}\ }\textbf {\bibinfo {volume} {75}},\ \bibinfo {pages} {4337} (\bibinfo {year} {1995})}\BibitemShut {NoStop}%
	\bibitem [{\citenamefont {Fabre}\ and\ \citenamefont {Treps}(2020)}]{FabreRev20}%
	\BibitemOpen
	\bibfield  {author} {\bibinfo {author} {\bibfnamefont {C.}~\bibnamefont {Fabre}}\ and\ \bibinfo {author} {\bibfnamefont {N.}~\bibnamefont {Treps}},\ }\bibfield  {title} {\bibinfo {title} {Modes and states in quantum optics},\ }\href {https://doi.org/10.1103/RevModPhys.92.035005} {\bibfield  {journal} {\bibinfo  {journal} {Rev. Mod. Phys.}\ }\textbf {\bibinfo {volume} {92}},\ \bibinfo {pages} {035005} (\bibinfo {year} {2020})}\BibitemShut {NoStop}%
	\bibitem [{\citenamefont {Sharmila}\ \emph {et~al.}(2020{\natexlab{a}})\citenamefont {Sharmila}, \citenamefont {Lakshmibala},\ and\ \citenamefont {Balakrishnan}}]{Sharmila20a}%
	\BibitemOpen
	\bibfield  {author} {\bibinfo {author} {\bibfnamefont {B.}~\bibnamefont {Sharmila}}, \bibinfo {author} {\bibfnamefont {S.}~\bibnamefont {Lakshmibala}},\ and\ \bibinfo {author} {\bibfnamefont {V.}~\bibnamefont {Balakrishnan}},\ }\bibfield  {title} {\bibinfo {title} {Signatures of avoided energy-level crossings in entanglement indicators obtained from quantum tomograms},\ }\href {https://doi.org/10.1088/1361-6455/abc07e} {\bibfield  {journal} {\bibinfo  {journal} {Journal of Physics B: Atomic, Molecular and Optical Physics}\ }\textbf {\bibinfo {volume} {53}},\ \bibinfo {pages} {245502} (\bibinfo {year} {2020}{\natexlab{a}})}\BibitemShut {NoStop}%
	\bibitem [{\citenamefont {Sharmila}\ \emph {et~al.}(2020{\natexlab{b}})\citenamefont {Sharmila}, \citenamefont {Lakshmibala},\ and\ \citenamefont {Balakrishnan}}]{Sharmila20b}%
	\BibitemOpen
	\bibfield  {author} {\bibinfo {author} {\bibfnamefont {B.}~\bibnamefont {Sharmila}}, \bibinfo {author} {\bibfnamefont {S.}~\bibnamefont {Lakshmibala}},\ and\ \bibinfo {author} {\bibfnamefont {V.}~\bibnamefont {Balakrishnan}},\ }\bibfield  {title} {\bibinfo {title} {Tomographic entanglement indicators in multipartite systems},\ }\href {https://doi.org/10.1007/s11128-020-02625-5} {\bibfield  {journal} {\bibinfo  {journal} {Quantum Information Processing}\ }\textbf {\bibinfo {volume} {19}},\ \bibinfo {pages} {127} (\bibinfo {year} {2020}{\natexlab{b}})}\BibitemShut {NoStop}%
	\bibitem [{\citenamefont {Sharmila}\ \emph {et~al.}(2019)\citenamefont {Sharmila}, \citenamefont {Lakshmibala},\ and\ \citenamefont {Balakrishnan}}]{Sharmila2019}%
	\BibitemOpen
	\bibfield  {author} {\bibinfo {author} {\bibfnamefont {B.}~\bibnamefont {Sharmila}}, \bibinfo {author} {\bibfnamefont {S.}~\bibnamefont {Lakshmibala}},\ and\ \bibinfo {author} {\bibfnamefont {V.}~\bibnamefont {Balakrishnan}},\ }\bibfield  {title} {\bibinfo {title} {Estimation of entanglement in bipartite systems directly from tomograms},\ }\href {https://doi.org/10.1007/s11128-019-2352-0} {\bibfield  {journal} {\bibinfo  {journal} {Quantum Information Processing}\ }\textbf {\bibinfo {volume} {18}},\ \bibinfo {pages} {236} (\bibinfo {year} {2019})}\BibitemShut {NoStop}%
	\bibitem [{\citenamefont {Sharmila}\ \emph {et~al.}(2017)\citenamefont {Sharmila}, \citenamefont {Saumitran}, \citenamefont {Lakshmibala},\ and\ \citenamefont {Balakrishnan}}]{Sharmila17}%
	\BibitemOpen
	\bibfield  {author} {\bibinfo {author} {\bibfnamefont {B.}~\bibnamefont {Sharmila}}, \bibinfo {author} {\bibfnamefont {K.}~\bibnamefont {Saumitran}}, \bibinfo {author} {\bibfnamefont {S.}~\bibnamefont {Lakshmibala}},\ and\ \bibinfo {author} {\bibfnamefont {V.}~\bibnamefont {Balakrishnan}},\ }\bibfield  {title} {\bibinfo {title} {Signatures of nonclassical effects in optical tomograms},\ }\href {https://doi.org/10.1088/1361-6455/aa51a4} {\bibfield  {journal} {\bibinfo  {journal} {Journal of Physics B: Atomic, Molecular and Optical Physics}\ }\textbf {\bibinfo {volume} {50}},\ \bibinfo {pages} {045501} (\bibinfo {year} {2017})}\BibitemShut {NoStop}%
	\bibitem [{\citenamefont {Hiltscher}\ \emph {et~al.}(2011)\citenamefont {Hiltscher}, \citenamefont {Governale}, \citenamefont {Splettstoesser},\ and\ \citenamefont {K\"onig}}]{hiltscher2011}%
	\BibitemOpen
	\bibfield  {author} {\bibinfo {author} {\bibfnamefont {B.}~\bibnamefont {Hiltscher}}, \bibinfo {author} {\bibfnamefont {M.}~\bibnamefont {Governale}}, \bibinfo {author} {\bibfnamefont {J.}~\bibnamefont {Splettstoesser}},\ and\ \bibinfo {author} {\bibfnamefont {J.}~\bibnamefont {K\"onig}},\ }\bibfield  {title} {\bibinfo {title} {Adiabatic pumping in a double-dot cooper-pair beam splitter},\ }\href@noop {} {\bibfield  {journal} {\bibinfo  {journal} {Phys. Rev. B}\ }\textbf {\bibinfo {volume} {84}},\ \bibinfo {pages} {155403} (\bibinfo {year} {2011})}\BibitemShut {NoStop}%
	\bibitem [{\citenamefont {Trocha}\ and\ \citenamefont {Weymann}(2015)}]{Trocha15}%
	\BibitemOpen
	\bibfield  {author} {\bibinfo {author} {\bibfnamefont {P.}~\bibnamefont {Trocha}}\ and\ \bibinfo {author} {\bibfnamefont {I.}~\bibnamefont {Weymann}},\ }\bibfield  {title} {\bibinfo {title} {Spin-resolved andreev transport through double-quantum-dot cooper pair splitters},\ }\href {https://doi.org/10.1103/PhysRevB.91.235424} {\bibfield  {journal} {\bibinfo  {journal} {Phys. Rev. B}\ }\textbf {\bibinfo {volume} {91}},\ \bibinfo {pages} {235424} (\bibinfo {year} {2015})}\BibitemShut {NoStop}%
	\bibitem [{\citenamefont {Souza}\ and\ \citenamefont {Sanz}(2017)}]{Souza17}%
	\BibitemOpen
	\bibfield  {author} {\bibinfo {author} {\bibfnamefont {F.~M.}\ \bibnamefont {Souza}}\ and\ \bibinfo {author} {\bibfnamefont {L.}~\bibnamefont {Sanz}},\ }\bibfield  {title} {\bibinfo {title} {Lindblad formalism based on fermion-to-qubit mapping for nonequilibrium open quantum systems},\ }\href {https://doi.org/10.1103/PhysRevA.96.052110} {\bibfield  {journal} {\bibinfo  {journal} {Phys. Rev. A}\ }\textbf {\bibinfo {volume} {96}},\ \bibinfo {pages} {052110} (\bibinfo {year} {2017})}\BibitemShut {NoStop}%
	\bibitem [{\citenamefont {Nogueira}\ \emph {et~al.}(2021)\citenamefont {Nogueira}, \citenamefont {Oliveira}, \citenamefont {Souza},\ and\ \citenamefont {Sanz}}]{Nogueira21}%
	\BibitemOpen
	\bibfield  {author} {\bibinfo {author} {\bibfnamefont {J.}~\bibnamefont {Nogueira}}, \bibinfo {author} {\bibfnamefont {P.~A.}\ \bibnamefont {Oliveira}}, \bibinfo {author} {\bibfnamefont {F.~M.}\ \bibnamefont {Souza}},\ and\ \bibinfo {author} {\bibfnamefont {L.}~\bibnamefont {Sanz}},\ }\bibfield  {title} {\bibinfo {title} {Dynamic generation of greenberger-horne-zeilinger states with coupled charge qubits},\ }\href {https://doi.org/10.1103/PhysRevA.103.032438} {\bibfield  {journal} {\bibinfo  {journal} {Phys. Rev. A}\ }\textbf {\bibinfo {volume} {103}},\ \bibinfo {pages} {032438} (\bibinfo {year} {2021})}\BibitemShut {NoStop}%
	\bibitem [{\citenamefont {Souza}\ \emph {et~al.}(2019)\citenamefont {Souza}, \citenamefont {Oliveira},\ and\ \citenamefont {Sanz}}]{Souza19}%
	\BibitemOpen
	\bibfield  {author} {\bibinfo {author} {\bibfnamefont {F.~M.}\ \bibnamefont {Souza}}, \bibinfo {author} {\bibfnamefont {P.~A.}\ \bibnamefont {Oliveira}},\ and\ \bibinfo {author} {\bibfnamefont {L.}~\bibnamefont {Sanz}},\ }\bibfield  {title} {\bibinfo {title} {Quantum entanglement driven by electron-vibrational mode coupling},\ }\href {https://doi.org/10.1103/PhysRevA.100.042309} {\bibfield  {journal} {\bibinfo  {journal} {Phys. Rev. A}\ }\textbf {\bibinfo {volume} {100}},\ \bibinfo {pages} {042309} (\bibinfo {year} {2019})}\BibitemShut {NoStop}%
	\bibitem [{\citenamefont {Oliveira}\ and\ \citenamefont {Sanz}(2015)}]{Oliveira15}%
	\BibitemOpen
	\bibfield  {author} {\bibinfo {author} {\bibfnamefont {P.}~\bibnamefont {Oliveira}}\ and\ \bibinfo {author} {\bibfnamefont {L.}~\bibnamefont {Sanz}},\ }\bibfield  {title} {\bibinfo {title} {Bell states and entanglement dynamics on two coupled quantum molecules},\ }\href@noop {} {\bibfield  {journal} {\bibinfo  {journal} {Annals of Physics}\ }\textbf {\bibinfo {volume} {356}},\ \bibinfo {pages} {244} (\bibinfo {year} {2015})}\BibitemShut {NoStop}%
	\bibitem [{\citenamefont {Nielsen}\ and\ \citenamefont {Chuang}(2010{\natexlab{a}})}]{nielsenbook}%
	\BibitemOpen
	\bibfield  {author} {\bibinfo {author} {\bibfnamefont {M.}~\bibnamefont {Nielsen}}\ and\ \bibinfo {author} {\bibfnamefont {I.~L.}\ \bibnamefont {Chuang}},\ }\href@noop {} {\emph {\bibinfo {title} {{Quantum Computation and Quantum Information}}}},\ \bibinfo {edition} {1st}\ ed.\ (\bibinfo  {publisher} {Cambridge University Press},\ \bibinfo {address} {Cambridge},\ \bibinfo {year} {2010})\BibitemShut {NoStop}%
	\bibitem [{\citenamefont {Nielsen}\ and\ \citenamefont {Chuang}(2010{\natexlab{b}})}]{niel}%
	\BibitemOpen
	\bibfield  {author} {\bibinfo {author} {\bibfnamefont {M.~A.}\ \bibnamefont {Nielsen}}\ and\ \bibinfo {author} {\bibfnamefont {I.~L.}\ \bibnamefont {Chuang}},\ }\href@noop {} {\emph {\bibinfo {title} {Quantum Computation and Quantum Information}}}\ (\bibinfo  {publisher} {Cambridge University Press, Cambridge},\ \bibinfo {year} {2010})\BibitemShut {NoStop}%
	\bibitem [{\citenamefont {\ifmmode~\dot{Z}\else \.{Z}\fi{}yczkowski}\ \emph {et~al.}(1998)\citenamefont {\ifmmode~\dot{Z}\else \.{Z}\fi{}yczkowski}, \citenamefont {Horodecki}, \citenamefont {Sanpera},\ and\ \citenamefont {Lewenstein}}]{pptHoro}%
	\BibitemOpen
	\bibfield  {author} {\bibinfo {author} {\bibfnamefont {K.}~\bibnamefont {\ifmmode~\dot{Z}\else \.{Z}\fi{}yczkowski}}, \bibinfo {author} {\bibfnamefont {P.}~\bibnamefont {Horodecki}}, \bibinfo {author} {\bibfnamefont {A.}~\bibnamefont {Sanpera}},\ and\ \bibinfo {author} {\bibfnamefont {M.}~\bibnamefont {Lewenstein}},\ }\bibfield  {title} {\bibinfo {title} {Volume of the set of separable states},\ }\href {https://doi.org/10.1103/PhysRevA.58.883} {\bibfield  {journal} {\bibinfo  {journal} {Phys. Rev. A}\ }\textbf {\bibinfo {volume} {58}},\ \bibinfo {pages} {883} (\bibinfo {year} {1998})}\BibitemShut {NoStop}%
	\bibitem [{\citenamefont {Sharmila}(2020)}]{sharmilathesis}%
	\BibitemOpen
	\bibfield  {author} {\bibinfo {author} {\bibfnamefont {B.}~\bibnamefont {Sharmila}},\ }\emph {\bibinfo {title} {Signatures of Nonclassical Effects in Tomograms}},\ \href@noop {} {\bibinfo {type} {Phd thesis}},\ \bibinfo  {school} {Indian Institute of Technology Madras}, \bibinfo {address} {Chennai, India} (\bibinfo {year} {2020}),\ \bibinfo {note} {available at \url{https://arxiv.org/abs/2009.09798}}\BibitemShut {NoStop}%
	\bibitem [{\citenamefont {Wootters}(1998)}]{Wootters98}%
	\BibitemOpen
	\bibfield  {author} {\bibinfo {author} {\bibfnamefont {W.~K.}\ \bibnamefont {Wootters}},\ }\bibfield  {title} {\bibinfo {title} {Entanglement of formation of an arbitrary state of two qubits},\ }\href@noop {} {\bibfield  {journal} {\bibinfo  {journal} {Phys. Rev. Lett.}\ }\textbf {\bibinfo {volume} {80}},\ \bibinfo {pages} {2245} (\bibinfo {year} {1998})}\BibitemShut {NoStop}%
	\bibitem [{\citenamefont {Hill}\ and\ \citenamefont {Wootters}(1997)}]{Hill97}%
	\BibitemOpen
	\bibfield  {author} {\bibinfo {author} {\bibfnamefont {S.}~\bibnamefont {Hill}}\ and\ \bibinfo {author} {\bibfnamefont {W.~K.}\ \bibnamefont {Wootters}},\ }\bibfield  {title} {\bibinfo {title} {Entanglement of a pair of quantum bits},\ }\href {https://doi.org/10.1103/PhysRevLett.78.5022} {\bibfield  {journal} {\bibinfo  {journal} {Phys. Rev. Lett.}\ }\textbf {\bibinfo {volume} {78}},\ \bibinfo {pages} {5022} (\bibinfo {year} {1997})}\BibitemShut {NoStop}%
	\bibitem [{\citenamefont {Elzerman}\ \emph {et~al.}(2004)\citenamefont {Elzerman}, \citenamefont {Hanson}, \citenamefont {Van~Beveren}, \citenamefont {Witkamp}, \citenamefont {Vandersypen},\ and\ \citenamefont {Kouwenhoven}}]{QDmeas}%
	\BibitemOpen
	\bibfield  {author} {\bibinfo {author} {\bibfnamefont {J.}~\bibnamefont {Elzerman}}, \bibinfo {author} {\bibfnamefont {R.}~\bibnamefont {Hanson}}, \bibinfo {author} {\bibfnamefont {L.~W.}\ \bibnamefont {Van~Beveren}}, \bibinfo {author} {\bibfnamefont {B.}~\bibnamefont {Witkamp}}, \bibinfo {author} {\bibfnamefont {L.}~\bibnamefont {Vandersypen}},\ and\ \bibinfo {author} {\bibfnamefont {L.~P.}\ \bibnamefont {Kouwenhoven}},\ }\bibfield  {title} {\bibinfo {title} {Single-shot read-out of an individual electron spin in a quantum dot},\ }\href@noop {} {\bibfield  {journal} {\bibinfo  {journal} {Nature}\ }\textbf {\bibinfo {volume} {430}},\ \bibinfo {pages} {431} (\bibinfo {year} {2004})}\BibitemShut {NoStop}%
	\bibitem [{\citenamefont {Breuer}\ and\ \citenamefont {Petruccione}(2007)}]{Breuerbook}%
	\BibitemOpen
	\bibfield  {author} {\bibinfo {author} {\bibfnamefont {H.-P.}\ \bibnamefont {Breuer}}\ and\ \bibinfo {author} {\bibfnamefont {F.}~\bibnamefont {Petruccione}},\ }\href@noop {} {\emph {\bibinfo {title} {{The Theory of Open Quantum Systems}}}},\ \bibinfo {edition} {1st}\ ed.\ (\bibinfo  {publisher} {Oxford University Press},\ \bibinfo {address} {Oxford},\ \bibinfo {year} {2007})\BibitemShut {NoStop}%
	\bibitem [{\citenamefont {Dehollain}\ \emph {et~al.}(2020)\citenamefont {Dehollain}, \citenamefont {Mukhopadhyay}, \citenamefont {Michal}, \citenamefont {Wang}, \citenamefont {Wunsch}, \citenamefont {Reichl}, \citenamefont {Wegscheider}, \citenamefont {Rudner}, \citenamefont {Demler},\ and\ \citenamefont {Vandersypen}}]{Dehollain20}%
	\BibitemOpen
	\bibfield  {author} {\bibinfo {author} {\bibfnamefont {J.~P.}\ \bibnamefont {Dehollain}}, \bibinfo {author} {\bibfnamefont {U.}~\bibnamefont {Mukhopadhyay}}, \bibinfo {author} {\bibfnamefont {V.~P.}\ \bibnamefont {Michal}}, \bibinfo {author} {\bibfnamefont {Y.}~\bibnamefont {Wang}}, \bibinfo {author} {\bibfnamefont {B.}~\bibnamefont {Wunsch}}, \bibinfo {author} {\bibfnamefont {C.}~\bibnamefont {Reichl}}, \bibinfo {author} {\bibfnamefont {W.}~\bibnamefont {Wegscheider}}, \bibinfo {author} {\bibfnamefont {M.~S.}\ \bibnamefont {Rudner}}, \bibinfo {author} {\bibfnamefont {E.}~\bibnamefont {Demler}},\ and\ \bibinfo {author} {\bibfnamefont {L.~M.~K.}\ \bibnamefont {Vandersypen}},\ }\bibfield  {title} {\bibinfo {title} {Nagaoka ferromagnetism observed in a quantum dot plaquette},\ }\href@noop {} {\bibfield  {journal} {\bibinfo  {journal} {Nature}\ }\textbf {\bibinfo {volume} {579}},\ \bibinfo {pages} {528} (\bibinfo {year} {2020})}\BibitemShut {NoStop}%
	\bibitem [{\citenamefont {Gustavsson}\ \emph {et~al.}(2006)\citenamefont {Gustavsson}, \citenamefont {Leturcq}, \citenamefont {Simovi\ifmmode~\check{c}\else \v{c}\fi{}}, \citenamefont {Schleser}, \citenamefont {Ihn}, \citenamefont {Studerus}, \citenamefont {Ensslin}, \citenamefont {Driscoll},\ and\ \citenamefont {Gossard}}]{Gustavsson06}%
	\BibitemOpen
	\bibfield  {author} {\bibinfo {author} {\bibfnamefont {S.}~\bibnamefont {Gustavsson}}, \bibinfo {author} {\bibfnamefont {R.}~\bibnamefont {Leturcq}}, \bibinfo {author} {\bibfnamefont {B.}~\bibnamefont {Simovi\ifmmode~\check{c}\else \v{c}\fi{}}}, \bibinfo {author} {\bibfnamefont {R.}~\bibnamefont {Schleser}}, \bibinfo {author} {\bibfnamefont {T.}~\bibnamefont {Ihn}}, \bibinfo {author} {\bibfnamefont {P.}~\bibnamefont {Studerus}}, \bibinfo {author} {\bibfnamefont {K.}~\bibnamefont {Ensslin}}, \bibinfo {author} {\bibfnamefont {D.~C.}\ \bibnamefont {Driscoll}},\ and\ \bibinfo {author} {\bibfnamefont {A.~C.}\ \bibnamefont {Gossard}},\ }\bibfield  {title} {\bibinfo {title} {Counting statistics of single electron transport in a quantum dot},\ }\href {https://doi.org/10.1103/PhysRevLett.96.076605} {\bibfield  {journal} {\bibinfo  {journal} {Phys. Rev. Lett.}\ }\textbf {\bibinfo {volume} {96}},\ \bibinfo {pages} {076605} (\bibinfo {year} {2006})}\BibitemShut {NoStop}%
	\bibitem [{\citenamefont {Fujisawa}\ \emph {et~al.}(2006)\citenamefont {Fujisawa}, \citenamefont {Hayashi},\ and\ \citenamefont {Sasaki}}]{Fujisawa_2006}%
	\BibitemOpen
	\bibfield  {author} {\bibinfo {author} {\bibfnamefont {T.}~\bibnamefont {Fujisawa}}, \bibinfo {author} {\bibfnamefont {T.}~\bibnamefont {Hayashi}},\ and\ \bibinfo {author} {\bibfnamefont {S.}~\bibnamefont {Sasaki}},\ }\bibfield  {title} {\bibinfo {title} {Time-dependent single-electron transport through quantum dots},\ }\href {https://doi.org/10.1088/0034-4885/69/3/R05} {\bibfield  {journal} {\bibinfo  {journal} {Reports on Progress in Physics}\ }\textbf {\bibinfo {volume} {69}},\ \bibinfo {pages} {759} (\bibinfo {year} {2006})}\BibitemShut {NoStop}%
	\bibitem [{\citenamefont {Shinkai}\ \emph {et~al.}(2009)\citenamefont {Shinkai}, \citenamefont {Hayashi}, \citenamefont {Ota},\ and\ \citenamefont {Fujisawa}}]{Shinkai09}%
	\BibitemOpen
	\bibfield  {author} {\bibinfo {author} {\bibfnamefont {G.}~\bibnamefont {Shinkai}}, \bibinfo {author} {\bibfnamefont {T.}~\bibnamefont {Hayashi}}, \bibinfo {author} {\bibfnamefont {T.}~\bibnamefont {Ota}},\ and\ \bibinfo {author} {\bibfnamefont {T.}~\bibnamefont {Fujisawa}},\ }\bibfield  {title} {\bibinfo {title} {Correlated coherent oscillations in coupled semiconductor charge qubits},\ }\href {https://doi.org/10.1103/PhysRevLett.103.056802} {\bibfield  {journal} {\bibinfo  {journal} {Phys. Rev. Lett.}\ }\textbf {\bibinfo {volume} {103}},\ \bibinfo {pages} {056802} (\bibinfo {year} {2009})}\BibitemShut {NoStop}%
	\bibitem [{\citenamefont {Fujisawa}\ \emph {et~al.}(2011)\citenamefont {Fujisawa}, \citenamefont {Shinkai}, \citenamefont {Hayashi},\ and\ \citenamefont {Ota}}]{Fujisawa11}%
	\BibitemOpen
	\bibfield  {author} {\bibinfo {author} {\bibfnamefont {T.}~\bibnamefont {Fujisawa}}, \bibinfo {author} {\bibfnamefont {G.}~\bibnamefont {Shinkai}}, \bibinfo {author} {\bibfnamefont {T.}~\bibnamefont {Hayashi}},\ and\ \bibinfo {author} {\bibfnamefont {T.}~\bibnamefont {Ota}},\ }\bibfield  {title} {\bibinfo {title} {Multiple two-qubit operations for a coupled semiconductor charge qubit},\ }\href {https://doi.org/https://doi.org/10.1016/j.physe.2010.07.040} {\bibfield  {journal} {\bibinfo  {journal} {Physica E: Low-dimensional Systems and Nanostructures}\ }\textbf {\bibinfo {volume} {43}},\ \bibinfo {pages} {730} (\bibinfo {year} {2011})},\ \bibinfo {note} {nanoPHYS 09}\BibitemShut {NoStop}%
	\bibitem [{\citenamefont {Hata}\ \emph {et~al.}(2023)\citenamefont {Hata}, \citenamefont {Sada}, \citenamefont {Uchino}, \citenamefont {Endo}, \citenamefont {Akiho}, \citenamefont {Muraki},\ and\ \citenamefont {Fujisawa}}]{Fujisawa23}%
	\BibitemOpen
	\bibfield  {author} {\bibinfo {author} {\bibfnamefont {T.}~\bibnamefont {Hata}}, \bibinfo {author} {\bibfnamefont {K.}~\bibnamefont {Sada}}, \bibinfo {author} {\bibfnamefont {T.}~\bibnamefont {Uchino}}, \bibinfo {author} {\bibfnamefont {D.}~\bibnamefont {Endo}}, \bibinfo {author} {\bibfnamefont {T.}~\bibnamefont {Akiho}}, \bibinfo {author} {\bibfnamefont {K.}~\bibnamefont {Muraki}},\ and\ \bibinfo {author} {\bibfnamefont {T.}~\bibnamefont {Fujisawa}},\ }\bibfield  {title} {\bibinfo {title} {Tunable tunnel coupling in a double quantum antidot with cotunneling via localized state},\ }\href {https://doi.org/10.1103/PhysRevB.108.075432} {\bibfield  {journal} {\bibinfo  {journal} {Phys. Rev. B}\ }\textbf {\bibinfo {volume} {108}},\ \bibinfo {pages} {075432} (\bibinfo {year} {2023})}\BibitemShut {NoStop}%
	\bibitem [{\citenamefont {Gramich}\ \emph {et~al.}(2017)\citenamefont {Gramich}, \citenamefont {Baumgartner},\ and\ \citenamefont {Sch\"onenberger}}]{Gramich17}%
	\BibitemOpen
	\bibfield  {author} {\bibinfo {author} {\bibfnamefont {J.}~\bibnamefont {Gramich}}, \bibinfo {author} {\bibfnamefont {A.}~\bibnamefont {Baumgartner}},\ and\ \bibinfo {author} {\bibfnamefont {C.}~\bibnamefont {Sch\"onenberger}},\ }\bibfield  {title} {\bibinfo {title} {Andreev bound states probed in three-terminal quantum dots},\ }\href {https://doi.org/10.1103/PhysRevB.96.195418} {\bibfield  {journal} {\bibinfo  {journal} {Phys. Rev. B}\ }\textbf {\bibinfo {volume} {96}},\ \bibinfo {pages} {195418} (\bibinfo {year} {2017})}\BibitemShut {NoStop}%
	\bibitem [{\citenamefont {Maisi}\ \emph {et~al.}(2011)\citenamefont {Maisi}, \citenamefont {Saira}, \citenamefont {Pashkin}, \citenamefont {Tsai}, \citenamefont {Averin},\ and\ \citenamefont {Pekola}}]{Maisi11}%
	\BibitemOpen
	\bibfield  {author} {\bibinfo {author} {\bibfnamefont {V.~F.}\ \bibnamefont {Maisi}}, \bibinfo {author} {\bibfnamefont {O.-P.}\ \bibnamefont {Saira}}, \bibinfo {author} {\bibfnamefont {Y.~A.}\ \bibnamefont {Pashkin}}, \bibinfo {author} {\bibfnamefont {J.~S.}\ \bibnamefont {Tsai}}, \bibinfo {author} {\bibfnamefont {D.~V.}\ \bibnamefont {Averin}},\ and\ \bibinfo {author} {\bibfnamefont {J.~P.}\ \bibnamefont {Pekola}},\ }\bibfield  {title} {\bibinfo {title} {Real-time observation of discrete andreev tunneling events},\ }\href {https://doi.org/10.1103/PhysRevLett.106.217003} {\bibfield  {journal} {\bibinfo  {journal} {Phys. Rev. Lett.}\ }\textbf {\bibinfo {volume} {106}},\ \bibinfo {pages} {217003} (\bibinfo {year} {2011})}\BibitemShut {NoStop}%
	\bibitem [{\citenamefont {Maisi}\ \emph {et~al.}(2014)\citenamefont {Maisi}, \citenamefont {Kambly}, \citenamefont {Flindt},\ and\ \citenamefont {Pekola}}]{Maisi14}%
	\BibitemOpen
	\bibfield  {author} {\bibinfo {author} {\bibfnamefont {V.~F.}\ \bibnamefont {Maisi}}, \bibinfo {author} {\bibfnamefont {D.}~\bibnamefont {Kambly}}, \bibinfo {author} {\bibfnamefont {C.}~\bibnamefont {Flindt}},\ and\ \bibinfo {author} {\bibfnamefont {J.~P.}\ \bibnamefont {Pekola}},\ }\bibfield  {title} {\bibinfo {title} {Full counting statistics of andreev tunneling},\ }\href {https://doi.org/10.1103/PhysRevLett.112.036801} {\bibfield  {journal} {\bibinfo  {journal} {Phys. Rev. Lett.}\ }\textbf {\bibinfo {volume} {112}},\ \bibinfo {pages} {036801} (\bibinfo {year} {2014})}\BibitemShut {NoStop}%
	\bibitem [{\citenamefont {Goldhaber-Gordon}\ \emph {et~al.}(1998)\citenamefont {Goldhaber-Gordon}, \citenamefont {Shtrikman}, \citenamefont {Mahalu}, \citenamefont {Abusch-Magder}, \citenamefont {Meirav},\ and\ \citenamefont {Kastner}}]{Goldhaber98}%
	\BibitemOpen
	\bibfield  {author} {\bibinfo {author} {\bibfnamefont {D.}~\bibnamefont {Goldhaber-Gordon}}, \bibinfo {author} {\bibfnamefont {H.}~\bibnamefont {Shtrikman}}, \bibinfo {author} {\bibfnamefont {D.}~\bibnamefont {Mahalu}}, \bibinfo {author} {\bibfnamefont {D.}~\bibnamefont {Abusch-Magder}}, \bibinfo {author} {\bibfnamefont {U.}~\bibnamefont {Meirav}},\ and\ \bibinfo {author} {\bibfnamefont {M.~A.}\ \bibnamefont {Kastner}},\ }\bibfield  {title} {\bibinfo {title} {Kondo effect in a single-electron transistor},\ }\href {https://doi.org/10.1038/34373} {\bibfield  {journal} {\bibinfo  {journal} {Nature}\ }\textbf {\bibinfo {volume} {391}},\ \bibinfo {pages} {156} (\bibinfo {year} {1998})}\BibitemShut {NoStop}%
	\bibitem [{\citenamefont {Horodecki}\ \emph {et~al.}(2009)\citenamefont {Horodecki}, \citenamefont {Horodecki}, \citenamefont {Horodecki},\ and\ \citenamefont {Horodecki}}]{HorodeckiRev}%
	\BibitemOpen
	\bibfield  {author} {\bibinfo {author} {\bibfnamefont {R.}~\bibnamefont {Horodecki}}, \bibinfo {author} {\bibfnamefont {P.}~\bibnamefont {Horodecki}}, \bibinfo {author} {\bibfnamefont {M.}~\bibnamefont {Horodecki}},\ and\ \bibinfo {author} {\bibfnamefont {K.}~\bibnamefont {Horodecki}},\ }\bibfield  {title} {\bibinfo {title} {Quantum entanglement},\ }\href {https://doi.org/10.1103/RevModPhys.81.865} {\bibfield  {journal} {\bibinfo  {journal} {Rev. Mod. Phys.}\ }\textbf {\bibinfo {volume} {81}},\ \bibinfo {pages} {865} (\bibinfo {year} {2009})}\BibitemShut {NoStop}%
\end{thebibliography}
\end{document}